\documentclass[showpacs,aps,prd,nofootinbib,floatfix,amsmath,amssymb]{revtex4}
\usepackage{graphicx}
 \usepackage{hyperref}

\newcommand{\al}			{\alpha}
\newcommand{\bt}			{\beta}

\newcommand{\dlt}			{\delta}
\newcommand{\eps}			{\epsilon}

\newcommand{\lm}			{\lambda}

\newcommand{\vfi}			{\varphi}
\newcommand{\ro}			{\rho}

\newcommand{\sgm}		{\sigma}

\newcommand{\Dlt}			{\Delta}

\newcommand{\mc}[1]{\mathcal{ #1}}						
\newcommand{\trm}[1]{\textrm{ #1}}						

\newcommand{\wt}[1]{\widetilde{ #1}}						

\newcommand{\ud}{\mathrm{d}} 								

\newcommand{\half}{\textstyle{\frac{1}{2}}}								
\newcommand{\quarter}{\textstyle{\frac{1}{4}}}							

\newcommand{\tbf}[1]{\textbf{#1}}														

\newcommand{\citte}[1]{~\cite{#1}}															

\newcommand{\ie}{\emph{i.e.\,}}															
\newcommand{\cf}{\emph{cf.\,}}															

\newcommand{\piw}{{\pi_{W}}}
\newcommand{\piy}{{\pi_{Y}}}
\newcommand{\piz}{{\pi_{Z}}}
\newcommand{\imc}[3]{\mc{#1}^{#2}_{#3}}

\newtheorem{proposition}{Proposition}[section]	
\numberwithin{equation}{section}				

\begin{document}
\makeatletter
\newbox\slashbox \setbox\slashbox=\hbox{$/$}
\newbox\Slashbox \setbox\Slashbox=\hbox{\large$/$}
\def\pFMslash#1{\setbox\@tempboxa=\hbox{$#1$}
  \@tempdima=0.5\wd\slashbox \advance\@tempdima 0.5\wd\@tempboxa
  \copy\slashbox \kern-\@tempdima \box\@tempboxa}
\def\pFMSlash#1{\setbox\@tempboxa=\hbox{$#1$}
  \@tempdima=0.5\wd\Slashbox \advance\@tempdima 0.5\wd\@tempboxa
  \copy\Slashbox \kern-\@tempdima \box\@tempboxa}
\def\FMslash{\protect\pFMslash}
\def\FMSlash{\protect\pFMSlash}
\def\miss#1{\ifmmode{/\mkern-11mu #1}\else{${/\mkern-11mu #1}$}\fi}
\makeatother

\title{Hidden symmetries induced by a canonical transformation and gauge structure of compactified Yang--Mills theories}
\author{M. A. L\' opez-Osorio$^{(a)}$, E. Mart\'\i nez-Pascual$^{(a)}$, H. Novales-S\' anchez$^{(b)}$, and J. J. Toscano$^{(a)}$}
\address{$^{(a)}$Facultad de Ciencias F\'{\i}sico Matem\'aticas,
Benem\'erita Universidad Aut\'onoma de Puebla, Apartado Postal
1152, Puebla, Puebla, M\'exico.\\
$^{(b)}$Facultad de Ciencias F\'\i sico Matem\' aticas,
Universidad Michoacana de San Nicol\' as de
Hidalgo, Avenida Francisco J. M\' ujica S/N, 58060, Morelia, Michoac\'an, M\' exico.}

\begin{abstract}
Compactified Yang-Mills theories with one universal extra dimension were found  [Phys.~Rev.~D \textbf{82}, 116012 (2010)] to exhibit two  types of gauge invariances: the standard gauge transformations (SGTs) and the nonstandard gauge transformations (NSGTs). In the present work we show that these transformations are not exclusive to compactified scenarios. Introducing a notion of hidden symmetry, based on the fundamental concept of canonical transformation, we analyse three different gauge systems, each of which is mapped to a certain effective theory that is invariant under the so-called SGTs and NSGTs. The systems under discussion are:  (i) four dimensional pure $ SU(3) $ Yang-Mills theory, (ii) four dimensional $ SU(3) $ Yang-Mills with spontaneous symmetry breaking, and (iii) pure Yang-Mills theory  with one universal compact extra dimension. The canonical transformation, that induces the notion of hidden symmetry, maps objects with well defined transformation laws under a gauge group $ G $ to well defined objects under a  non-trivial subgroup $ H\subset G $. In the case where spontaneous symmetry breaking is present, the set of SGTs corresponds to the group into which the original gauge group is broken into, whereas the NSGTs are associated to the broken generators and can be used to define the unitary gauge. For the system (iii), the SGTs coincide with the gauge group $ SU(N,\mc{M}^{4}) $, whereas the NSGTs do not form a group; in this system the `fundamental' theory and the effective one are shown to be classically equivalent. 
\end{abstract}


\maketitle

\section{Introduction}
\label{int}

Recently, the ATLAS~\cite{ATLAS} and CMS~\cite{CMS} experiments at the Large Hadron Collider reported the presence of a scalar boson with mass in the range $125-126 \, GeV$ that is compatible with the Standard Model Higgs boson. If couplings of this particle to pairs of  $W$ and $Z$ weak gauge bosons are found to coincide with those predicted by the Standard Model in subsequent analysis of experimental data, the Higgs mechanism~\cite{HM} will be firmly established as a genuine phenomenon of nature. Since the Higgs mechanism endows gauge bosons with mass through absorption of Goldstone bosons~\cite{GT}, resulting in spontaneous symmetry breaking (SSB), experimental data would confirm that the weak interaction is spontaneously broken. This in turn would validate the existence of a degenerate vacuum as the source of elementary particle masses and constitutes a good motivation to investigate new mechanisms of mass generation. In particular, it is interesting to study some kind of source, alternative to spontaneous symmetry breaking, that still allows the Higgs mechanism to operate. In this direction, it is already known that gauge theories formulated on spacetime manifolds with compact extra dimensions~\cite{EDA} enable endowing with mass the Kaluza-Klein gauge excitations in the absence of  degenerate vacuum. Although in these theories there are pseudo-Goldstone bosons,  thus allowing the Higgs mechanism to operate, they do not correspond to genuine Goldstone bosons in the sense of spontaneous breakdown of a global symmetry. The emergent pseudo-Goldstone bosons in Kaluza-Klein theories are directly generated  by compactification of the spatial extra dimensions.

In this work we clarify the gauge structure of pure Yang-Mills theories formulated on spacetime manifolds with compact spatial extra dimensions through a novel notion of hidden symmetry. Some theoretical aspects of these theories have already been studied in Refs.~\cite{NT,NT2,NT3,EDT}. Also, they have been the subject of important phenomenological interest in the context of dark matter~\cite{UEDDM}, neutrino physics~\cite{UEDn}, Higgs physics~\cite{UEDH}, flavor physics~\cite{UEDf}, hadronic and linear colliders~\cite{UEDc}, and electroweak gauge couplings~\cite{NT4}. In Ref.~\cite{NT}, some results in the context of a pure Yang-Mills theory with one universal extra dimension (UED) were presented. For instance, the necessity of explaining the gauge structure of the compactified theory in order to quantize it was emphasized. In particular, it was noticed that as a consequence of compactification, the original gauge transformations split into two classes: the standard gauge transformations (SGTs) and the nonstandard gauge transformations (NSGTs). In the following pages these transformations are formulated, via a certain canonical transformation, within the framework of hidden symmetry. 

The concept of hidden symmetry is usually associated to theories in which SSB occurs; however,  we will show that this is not exclusive to theories of this kind. A symmetry, encoded in a gauge group $ G $, can also be hidden if  there is a canonical transformation that  maps well defined objects under the group $G$ to well defined objects under a subgroup $H\subset G$. As we will show below, this transformation is crucial to understanding a hidden symmetry in this context. It is at the level of the remaining symmetry that SGTs and NSGTs find a clear interpretation. The set of SGTs forms a group which coincides with the subgroup $ H $, whereas the set of NSGTs does not form a group, as will be shown. Nevertheless, the phenomenon of SSB can be fit into this general scenario.  When a scalar sector that leads to a degenerate vacuum which is invariant under the subgroup $H$ is introduced, the NSGTs are associated to the broken generators of the  group $G$.

The novel point of view of hidden symmetry given in this paper will be exemplified by studying in detail three gauge models. The first of them shows that our notion of hidden symmetry is not necessarily embedded either in a compactification scheme from a higher dimensional theory or in a theory that presents SSB. Using the four dimensional pure $ SU(3) $ Yang-Mills theory, we explicitly construct a canonical transformation which maps gauge fields of $ SU(3) $ into gauge fields, two doublets and a singlet with respect to the subgroup $ SU(2) $. In this case, the $ SU(3) $ symmetry is hidden into the SGTs and NSGTs, which correspond to transformations in $ SU(2) $ and transformations related with the five remainder generators of $ SU(3) $, respectively. 

The second model under consideration is an $ SU(3) $ Yang-Mills theory with a renormalizable scalar sector that presents SSB; our canonical transformation will decompose  the $ SU(3) $ Yang-Mills connection and the matter scalar into well defined objects with respect to $ SU(2) $. The new ingredients in the analysis of this model are twofold. On one hand, these will help us to compare a hidden symmetry arising from SSB  with a hidden symmetry in terms of the suitable canonical transformation. On the other hand, they will provide a way  to clarify the physical meaning of the NSGTs by showing that the unitary gauge corresponds to a particular transformation of this type.

The third system on which we focus our attention is an effective theory that results from compactifying the spatial extra dimension of a manifold $\mc{M}^{5}$ on which a pure Yang-Mills theory is defined. The compactification scheme is achieved along the same lines of Ref.~\cite{NT}. The basic fields of the higher dimensional theory are gauge fields under the gauge group $ SU(N,\mc{M}^{5}) $--that is, the group $ SU(N) $ with gauge parameters propagating in the bulk $ \mc{M}^{5} $. Defined on the Minkowski spacetime $ \mc{M}^{4} $, the effective theory existing alongside  carries the four-dimensional Fourier modes of the five-dimensional gauge fields as basic fields; we show that Fourier expansions in this case determine a canonical transformation which maps well defined objects under $ SU(N,\mc{M}^{5}) $ to well defined objects under $ SU(N,\mc{M}^{4}) $. In our terminology, the gauge symmetry of the higher-dimensional theory is just hidden in the lower-dimensional theory --that is,  $ SU(N,\mc{M}^{5}) $ is codified in the SGTs and NSGTs, the former being represented by $ SU(N,\mc{M}^{4}) $. It is worth noticing that in the effective theory there emerge a massless scalar bosons which can be removed by a specific NSGT, these correspond to  pseudo-Goldstone bosons which remarkably do not arise from SSB mechanism; in this scenario, compactification does not involve broken generators; it entails a change in the support manifold of the group parameters. The specific NSGT to eliminate the pseudo-Goldstone bosons can hence be interpreted as a unitary gauge~\cite{NT}. 

The suitable canonical transformations found in these systems, through which the original symmetry is hidden, permeates at the level of the Dirac algorithm for constrained systems. In each case, the primary Hamiltonian and each generation of constraints, of the theory manifestly invariant under the group $ G  $, are mapped onto the corresponding primary Hamiltonian and generation of constraints, present in the theory invariant under SGTs and NSGTs. Since every model we analyzed is a first-class constraint system, this implies that the canonical transformation for each case maps the gauge generator of the group $ G $ onto the gauge generators of the SGTs and NSGTs. This result is particularly interesting for the third system, as it implies that the gauge structure of the higher-dimensional theory is certainly rewritten in terms of SGTs and NSGTs; by counting degrees of freedom, we will show that the five-dimensional and the effective theory are equivalent at the classical level.

The rest of the paper has been organized as follows: In Sec.~\ref{TM}, the pure $ SU(3) $ Yang-Mills theory is introduced, and canonical analysis of the theory both before and after considering a suitable canonical transformation are independently achieved. It is shown that both frameworks lead to the same theory with the same number of physical degrees of freedom and the same gauge transformations; \ie, the canonical transformation simply recasts the system. In Sec.~\ref{SU3SSB} a renormalizable scalar Higgs sector is added to the model presented in Sec. \ref{TM}, and the corresponding suitable canonical transformation is introduced. We show that the presence of  spontaneous breakdown $SU(3) \to SU(2)$ allows us to use a specific NSGT as the unitary gauge.  Section \ref{YM5} is devoted to the study of pure $ SU(N,\mc{M}) $ Yang-Mills theory in an arbitrary number of dimensions; with the more tractable case of one UED, we explicitly present the suitable canonical transformation and compactification scheme that led us to the effective theory invariant under SGTs and NSGTs. We argue that both theories are equivalent as they have the same gauge transformations, simply written in different coordinates, and contain the same number of physical degrees of freedom. In Sec.~\ref{FR}, a summary of our results is presented. Finally, in the Appendix, we collect the proof on the canonical nature of the Fourier transform.

\section{The toy model: Pure {$SU(3)$} Yang-Mills theory}
\label{TM}
The purpose of this section is to illustrate the notion of hidden symmetry within the context described in the Introduction, for which we consider the situation where $G=SU(3)$ and $H=SU(2)$, and construct the desirable canonical transformation. This system neither presents SSB nor is involved in any compactification scheme. This model has attracted important phenomenological interest within the context of the so-called 331 models~\cite{331}, and it is useful for us because the $SU(2)$ group is completely embedded in the $SU(3)$ one. This feature allows us to clearly illustrate all the peculiarities of the notion of hidden symmetry that we are introducing in this paper.

\subsection{The {$SU(3)$} perspective of the model}\label{secYM3}
We consider the four-dimensional Yang-Mills theory based on the $ SU(3) $ group with the well-known Lagrangian
\begin{equation}
\label{L3}
{\mc L}_{SU(3)}=-\frac{1}{4}F^a_{\mu \nu}F^{\mu \nu}_a \ ,
\end{equation}
where the components of the Yang-Mills curvature are given in terms of the gauge fields $ A^{a}_{\mu} $ by
\begin{equation}\label{fmn3}
F^a_{\mu \nu}=\partial_\mu A^a_\nu-\partial_\nu A^a_\mu +gf^{abc}A^b_\mu A^c_\nu \ .
\end{equation}
In the special case of $ SU(3) $, the completely antisymmetric structure constants $ f^{abc} $ have the following nonvanishing values: $ f^{123}=1 $, $ f^{147}=-f^{156}=f^{246}=f^{257}=f^{345}=-f^{367}=\half $ and $ f^{458}=f^{678}=\textstyle{\frac{\sqrt{3}}{2}} $.

The Lagrangian \eqref{L3} is invariant under gauge transformations
\begin{equation}
\label{gt}
\delta A^a_\mu(x)={\mc D}^{ab}_\mu \alpha^b(x)\ ,
\end{equation}
where $\alpha^a$ are the gauge parameters of the group and ${\mc D}^{ab}_\mu=\delta^{ab}\partial_\mu -gf^{abc}A^c_\mu$ is the covariant derivative in the adjoint representation. The above gauge transformations imply that the components of the curvature transform in the adjoint representation of the group,
\begin{equation}
\label{tc}
\delta F^a_{\mu \nu}=gf^{abc}F^b_{\mu \nu}\alpha^c \ .
\end{equation}

As far as the Hamiltonian structure of the theory is concerned, the canonical momenta are defined by
\begin{equation}\label{pis3}
\pi^\mu_a \equiv\frac{\partial {\mc L}_{SU(3)}}{\partial \dot{A}^{a}_{\mu}} =F_a^{\mu 0}\ ,
\end{equation}
where the dot over the fields denotes time derivative. This expression immediately leads to the following primary constraints:
\begin{equation}
\label{pc3}
\phi^{(1)}_a\equiv \pi_a^0 \approx 0 \ ,
\end{equation}
where the symbol $ \approx $ denotes weakly zero~\cite{dirbk64}. The time evolution along the motion of an arbitrary function on the phase space is dictated by the primary Hamiltonian
\begin{equation}\label{phsu3}
\mc{H}^{(1)}_{SU(3)}={\mc H}_{SU(3)}+\mu^a \phi^{(1)}_a \  ,
\end{equation}
where $\mu^a$ are Lagrange multipliers, and ${\mc H}_{SU(3)}$ is the canonical Hamiltonian. The latter is
\begin{equation}
\label{ch3}
{\mc H}_{SU(3)}=\frac{1}{2}\pi_a^i \pi_a^i+\frac{1}{4}F^a_{ij}F^{ij}_a-A_0^a\phi^{(2)}_{a} \ .
\end{equation}

Any physically allowed initial configuration of fields and conjugate momenta must satisfy the primary constraints \eqref{pc3}; hence the constraints must be constant in time. This consistency condition on the primary constraints leads to the following secondary constraints:
\begin{equation}
\label{sc3}
\phi^{(2)}_a\equiv {\mc D}^{ab}_i \pi_b^i \approx 0\, .
\end{equation}
Applying the consistency condition to secondary constraints yields no new constraints. In this case, all the constraints are of first-class type \cite{dirbk64}; the Poisson brackets among the constraints are linear combinations of the constraints themselves. The nonvanishing Poisson brackets between first-class constraints are
\begin{equation}
\label{la3}
\{ \phi^{(2)}_a[u],\phi^{(2)}_b[v]\}_{SU(3)}=gf_{abc}\,\phi^{(2)}_c[uv]\ ,
\end{equation}
where smeared form $ \,\phi^{(2)}_a[u]:=\int d^{3}x\, u(\mathbf{x}) \phi^{(2)}_a(\mathbf{x})$  of the constraints was used. The label $ SU(3) $ on the Poisson bracket indicates that it is calculated with respect to the canonical conjugate pairs $ (A_{\mu}^{a},\pi_{a}^{\mu}) $.

As is well known \cite{henbk} the number of true degrees of freedom, in a theory with first-class constraints only, corresponds to the total number of canonical variables minus twice the number of first-class constraints, all divided by two. Therefore, the number of true degrees of freedom is in this case $ 16$ per spatial point $ \textbf{x} $ of Minkowski spacetime $ \mc{M}^{4} $.

In this system all first-class constraints generate gauge transformations \eqref{gt} through the gauge generator \cite{cas82}
\begin{equation}\label{gg3}
G= (\mc{D}^{ab}_{0}\al^{b})\phi^{(1)}_{a}-\al^{a}\phi^{(2)}_{a}
\end{equation}
via the Poisson bracket as follows
\begin{equation}\label{gtcasu3}
\dlt A_{\mu}^{a}= \{A_{\mu}^{a},G\}_{SU(3)}\ .
\end{equation}

We now turn to formulate the same theory but from the perspective of $SU(2)$.

\subsection{The {$SU(2)$} perspective of the model}\label{SU2}
The fundamental representation of $ SU(3) $ has dimension $3$. A particular choice of this representation is given by the well known Gell-mann matrices $\lambda^a$, being the corresponding generators $\lambda^a/2$. Since matrices $\lambda^3$ and $\lambda^8$ commute with each other, there are three independent $SU(2)$ subgroups, whose generators are $(\lambda^1,\lambda^2,\lambda)$, $(\lambda^4,\lambda^5,\lambda)$ and $(\lambda^6,\lambda^7,\lambda)$. For each case, $\lambda$ is a different linear combination (with real coefficients) of $\lambda^3$ and $\lambda^8$. In this work, we will consider the subgroup determined by the set of generators $(\lambda^1,\lambda^2,\lambda^3)$ and the corresponding values of the structure constants will be denoted by $ f^{\bar{a}\bar{b}\bar{c}}=\eps^{\bar{a}\bar{b}\bar{c}} $, where $ \bar{a}=1,2,3 $. We will also use the notation $ \hat{a}=4,5,6,7 $ so that $ a=1,\ldots,8=\bar{a},\hat{a},8 $. From the $ SU(2) $ perspective $ \bar{a} $ will label gauge fields, whereas $ \hat{a} $ and $ 8 $ will label tensorial representations of $ SU(2) $, see Eqs. \eqref{sgt2}.

In the configuration space, we consider the following point transformation:
\begin{subequations}\label{pt1}
\begin{align}
A^{\bar{a}}_\mu & = W^{\bar{a}}_\mu  \ , \label{ptbara}\\
A^{4}_\mu & = \frac{1}{\sqrt{2}} \left(Y_{\mu 1}^{*}+Y_{\mu}^{1}\right)\,  ,\ A^{5}_\mu = \frac{-i}{\sqrt{2}} \left(Y_{\mu 1}^{*}-Y_{\mu}^{1}\right) \, , \label{pt45}\\
A^{6}_\mu & = \frac{1}{\sqrt{2}} \left(Y_{\mu 2}^{*}+Y_{\mu}^{2}\right)\,  ,\ A^{7}_\mu = \frac{-i}{\sqrt{2}} \left(Y_{\mu 2}^{*}-Y_{\mu}^{2}\right)\,  ,  \label{pt67}\\
A^{8}_\mu & = Z_{\mu}\  .  \label{pt8}
\end{align}
\end{subequations}
This mapping relates the coordinates of the $ SU(3) $ formulation to the coordinates we will use in the $ SU(2) $ perspective. The inverse is conveniently arranged as follows:
\begin{subequations}\label{invpt1}
\begin{align}
W^{\bar{a}}_\mu & =  A^{\bar{a}}_\mu\ , \label{ptW}\\
Y_{\mu} & =
\begin{pmatrix}
Y_{\mu}^{1}\\
Y_{\mu}^{2}
\end{pmatrix} = \frac{1}{\sqrt{2}}
\begin{pmatrix}
A_{\mu}^{4}-iA_{\mu}^{5}\\
A_{\mu}^{6}-iA_{\mu}^{7}
\end{pmatrix}\ , \label{ptY}\\
Y^{\dagger}_{\mu} & =(Y_{\mu\,1}^{*}\ \ Y_{\mu\,2}^{*})=\dfrac{1}{\sqrt{2}}(A_{\mu}^{4}+iA_{\mu}^{5}\ \  A_{\mu}^{6}+iA_{\mu}^{7})\ ,\label{ptYstar}\\
Z_{\mu} & =A_{\mu}^{8}\ . \label{ptZ}
\end{align}
\end{subequations}
As will be confirmed below, see Eq.\eqref{sgt2}, fields $ Y_{\mu} $ and  $ Y_{\mu}^{\dag} $ transform as contravariant and covariant $ SU(2) $ objects, respectively, whereas $ Z_{\mu} $ becomes invariant under this group.

Using the above point transformation, the Yang-Mills curvature components \eqref{fmn3} can be rearranged as follows:
\begin{subequations}\label{fmn2}
\begin{align}
F^{\bar{a}}_{\mu \nu} & = W^{\bar{a}}_{\mu \nu}+ig\left(Y^\dag_\mu \frac{\sigma^{\bar{a}}}{2}Y_\nu-Y^\dag_\nu \frac{\sigma^{\bar{a}}}{2}Y_\mu \right)\ , \label{fmnbara}\\
Y_{\mu\nu} & = D_\mu Y_\nu-D_\nu Y_\mu +ig \frac{\sqrt{3}}{2}\left(Y_\mu Z_\nu-Y_\nu Z_\mu\right) \ , \label{ymn}\\
F_{\mu\nu}^{8} & = Z_{\mu \nu}+ig \frac{\sqrt{3}}{2}\left(Y^\dag_\mu Y_\nu-Y^\dag_\nu Y_\mu \right)\ .\label{fmn8}
\end{align}
\end{subequations}
In these equations, $W^{\bar{a}}_{\mu \nu}=\partial_\mu W^{\bar{a}}_\nu-\partial_\nu W^{\bar{a}}_\mu+g\epsilon^{\bar{a}\bar{b}\bar{c}}\,W^{\bar{b}}_\mu W^{\bar{c}}_\nu$ are the components of the $su(2)$-valued curvature,  $D_\mu=\partial_\mu -ig\frac{\sigma^{\bar{a}}}{2}W^{\bar{a}}_\mu$ is the covariant derivative in the fundamental representation of $SU(2)$, and $Z_{\mu \nu}=\partial_\mu Z_\nu-\partial_\nu Z_\mu$. The components $ F_{\mu\nu}^{\hat{a}} $ are encoded into $ Y_{\mu\nu} $.

In terms of  $ SU(2) $ objects, the Lagrangian \eqref{L3} becomes
\begin{equation}\label{L2}
{\mc L}_{SU(2)}=-\frac{1}{4}F^{\bar{a}}_{\mu \nu}F^{\mu \nu}_{\bar{a}}-\frac{1}{2}Y^\dag_{\mu \nu}Y^{\mu \nu}-\frac{1}{4}F^8_{\mu \nu}F^{\mu \nu}_8 \ ,
\end{equation}
and the gauge transformations (\ref{gt}) are mapped onto
\begin{subequations}\label{gtsu2}
\begin{align}
\delta W^{\bar{a}}_\mu & = {\mc D}^{\bar{a}\bar{b}}_\mu \alpha^{\bar{b}}-ig\left(\beta^\dag \frac{\sigma^{\bar{a}}}{2}Y_\mu -Y^\dag_\mu \frac{\sigma^{\bar{a}}}{2} \beta \right)\ , \label{gt1}\\
\delta Y_\mu &=ig \frac{\sigma^{\bar{a}}}{2}\alpha^{\bar{a}}\, Y_\mu +\left(D_\mu -ig \frac{\sqrt{3}}{2}Z_\mu\right)\beta +ig \frac{\sqrt{3}}{2}\, Y_\mu \alpha_Z\ , \label{gt2}\\
\delta Z_\mu &=\partial_\mu \alpha_Z-ig \frac{\sqrt{3}}{2}\left(\beta^\dag Y_\mu-Y^\dag_\mu \beta\right) \ ,\label{bt3}
\end{align}
\end{subequations}
where $\beta^\dag =\left(\frac{1}{\sqrt{2}}\left(\alpha^4+i\alpha^5\right)\ \ \frac{1}{\sqrt{2}}\left(\alpha^6+i\alpha^7\right)\right)$. From the $SU(2)$ perspective, the eight parameters of $SU(3)$ split into three gauge parameters, $\alpha^{\bar{a}}$, two doublets, $\beta$ and $ \bt^{\dag} $, and a singlet, $\alpha_Z$, of $SU(2)$. In Eq.~\eqref{gt1} the covariant derivative of $ SU(2) $, ${\mc D}^{\bar{a}\bar{b}}_\mu=\delta^{\bar{a}\bar{b}}\partial_\mu -g\epsilon^{\bar{a}\bar{b}\bar{c}}W^{\bar{c}}_\mu$, in its adjoint representation, emerges.

The \emph{standard gauge transformations} (SGTs) are defined from the transformation laws Eq.~\eqref{gtsu2} by setting the parameters $\beta$ and $\alpha_Z$ equal to zero,
\begin{subequations}\label{sgt2}
\begin{align}
\delta_{\!\!\trm{s}} W^{\bar{a}}_\mu & \equiv {\mc D}^{\bar{a}\bar{b}}_\mu \alpha^{\bar{b}}\  ,\label{sgtW}\\
\delta_{\!\!\trm{s}} Y_\mu & \equiv ig \frac{\sigma^{\bar{a}}}{2}\alpha^{\bar{a}}\, Y_\mu \ , \label{sgtY}\\
\delta_{\!\!\trm{s}} Z_\mu & \equiv 0\ .\label{sgtZ}
\end{align}
\end{subequations}
From these expressions, it is shown $W^{\bar{a}}_\mu$ transform as gauge fields, $Y_{\mu}$  as a doublet of $SU(2)$, and $Z_\mu$ as a scalar under $ SU(2) $. This means that the transformation defined by Eq.~\eqref{pt1} constitutes an admissible point transformation, as well defined objects under $SU(3)$ are mapped onto well defined objects under $SU(2)$. Moreover, Eqs.~\eqref{sgtY} and \eqref{sgtZ}  make manifest that $Y_\mu$ and $Z_\mu$ are matter fields.  In the context of this description, there arise \emph{nonstandard gauge transformations} (NSGTs), which are defined from Eq.~\eqref{gtsu2} by setting $\alpha^{\bar{a}}=0$:
\begin{subequations}\label{nsgtsu2}
\begin{align}
\delta_{\!\!\trm{ns}} W^{\bar{a}}_\mu &= -ig\left(\beta^\dag \frac{\sigma^{\bar{a}}}{2}Y_\mu -Y^\dag_\mu \frac{\sigma^{\bar{a}}}{2} \beta \right)\ ,\label{nsgt1}\\
\delta_{\!\!\trm{ns}} Y_\mu &=\left(D_\mu -ig \frac{\sqrt{3}}{2}Z_\mu\right)\beta +ig \frac{\sqrt{3}}{2}\, Y_\mu \alpha_Z\ ,\label{nsgt2}\\
 \delta_{\!\!\trm{ns}} Z_\mu &=\partial_\mu \alpha_Z-ig \frac{\sqrt{3}}{2}\left(\beta^\dag Y_\mu-Y^\dag_\mu \beta\right) \ ,\label{nsgt3}
\end{align}
\end{subequations}
These NSGT tell us that there is a gauge symmetry larger than $SU(2)$, in our case $SU(3)$. \textit{More precisely, the difference between SGTs and NSGTs in this model is that the former are associated with generators that constitute a group, whereas the latter have to do with generators that do not share this property}. We will discuss further on this point at the end of this section. {This differentiation is crucial to quantizing the theory, as it requires incorporating the gauge parameters as degrees of freedom from the beginning, so the use of only the $SU(2)$ parameters or the complete set of the $SU(3)$ parameters would lead to very different quantized theories.} Of course, the theory can be quantized using the $SU(2)$ basis but taking into account that $Y_\mu$ and $Z_\mu$  are also gauge fields, which means that the $\beta$ and $\alpha_Z$ parameters must be recognized as (ghost) degrees of freedom in the context of the  BRST symmetry~\cite{BRST,AFAB}.

In order to justify that the components of Eq.~\eqref{gtsu2} are actual gauge transformations, we wish to show the invariance of the Lagrangian Eq.~\eqref{L2} under these variations. Therefore, one may want to start exploring the behaviour of Eq.~\eqref{fmn2} under Eq.~\eqref{gtsu2}. After some algebra, one finds
\begin{subequations}\label{gtfmn2}
\begin{align}
\delta F^{\bar{a}}_{\mu \nu} & = g\epsilon^{\bar{a}\bar{b}\bar{c}}F^{\bar{b}}_{\mu \nu}\alpha^{\bar{c}}+ig\left(Y^\dag_{\mu \nu}\frac{\sigma^{\bar{a}}}{2}\beta-\beta^\dag\frac{\sigma^{\bar{a}}}{2}Y_{\mu \nu}  \right)\ ,\label{gtfmnbara}\\
\delta Y_{\mu \nu} & = ig\frac{\sigma^{\bar{a}}}{2}Y_{\mu \nu}\alpha^{\bar{a}}-igF^{\bar{a}}_{\mu \nu}\frac{\sigma^{\bar{a}}}{2}\beta +ig\frac{\sqrt{3}}{2}\left( Y_{\mu \nu}\alpha_Z-F^8_{\mu \nu}\beta\right)\ ,\label{gtYmn}\\
\delta F^8_{\mu \nu}&=ig\frac{\sqrt{3}}{2}\left(Y^\dag_{\mu \nu}\beta-\beta^\dag Y_{\mu \nu}\right) \ .\label{gtfmn8}
\end{align}
\end{subequations}
It can be shown that the Lagrangian in the $ SU(2) $ description Eq.~\eqref{L2} is invariant under these transformations. Therefore it is also invariant under the transformations in Eq.~\eqref{gtsu2} that decompose into the sum of SGTs and NSGTs.

We now place some technical comments. In the variation of $ Y_{\mu\nu} $~Eq.~\eqref{gtYmn}, the following extra term is explicitly found
\begin{align*}
B_{\mu\nu}  =& -g^{2}\Big[\big(Y_{\mu}^{\dag}\frac{\sgm^{\bar{a}}}{2}Y_{\nu}-Y_{\nu}^{\dag}\frac{\sgm^{\bar{a}}}{2}Y_{\mu}\big)\frac{\sgm^{\bar{a}}}{2}\bt+\frac{3}{4}\big(Y^{\dag}_{\mu}Y_{\nu}-Y^{\dag}_{\nu}Y_{\mu}\big)\bt+\frac{3}{4}\big(\bt^{\dag}Y_{\mu}-Y^{\dag}_{\mu}\bt\big)Y_{\nu}\\
 &  -\frac{3}{4}\big(\bt^{\dag}Y_{\nu}-Y^{\dag}_{\nu}\bt\big)Y_{\mu}+\big(\bt^{\dag}\frac{\sgm^{\bar{a}}}{2}Y_{\mu}-Y_{\mu}^{\dag}\frac{\sgm^{\bar{a}}}{2}\bt\big)\frac{\sgm^{\bar{a}}}{2}Y_{\nu}-\big(\bt^{\dag}\frac{\sgm^{\bar{a}}}{2}Y_{\nu}-Y_{\nu}^{\dag}\frac{\sgm^{\bar{a}}}{2}\bt\big)\frac{\sgm^{\bar{a}}}{2}Y_{\mu}\Big]\ .
\end{align*}
It at first sight seems to be different from zero, but consistency between the $ SU(2) $ and the $SU(3)$ perspectives of the same theory indicates that it must vanish. Indeed, using the point transformation Eq.~\eqref{invpt1}, the Eqs.~\eqref{gtfmnbara} to  \eqref{gtfmn8} can be rearranged to Eq.~\eqref{tc} as required. Also, since $ B_{\mu\nu} $ is linear in $ \bt $, its occurrence in the variation of $ Y_{\mu\nu} $ would spoil the invariance of the Lagrangian Eq.~\eqref{L2} under the NSGTs Eq.~\eqref{nsgtsu2}; however, one can see that the $ r $th $ SU(2) $ component ($ r=1,2 $) of the doublet $B_{\mu\nu}$ is of the form
\begin{equation*}
B_{\mu\nu}^{r}= -\frac{1}{4}\Big[\big(T^{rs}_{pq}-T^{rs}_{qp}\big)\big(Y_{\mu s}^{*}Y_{\nu}^{\;q}-Y_{\nu s}^{*}Y_{\mu}^{\;q}\big)\bt^{p}-T^{rs}_{pq}\big(Y_{\mu}^{\;p}Y_{\nu}^{\;q}-Y_{\nu}^{\;p}Y_{\mu}^{\;q}\big)\bt_{s}\Big]\ ,
\end{equation*}
where $ T^{rs}_{pq}\equiv (\sgm^{\bar{a}})^{r}_{p}(\sgm^{\bar{a}})^{s}_{q}+3\dlt^{r}_{p}\dlt^{s}_{q} $. Using the explicit values of the indices shows that $  T^{rs}_{pq} $ is symmetric in $ p $ and $ q$, hence $ B_{\mu\nu}^{r}\equiv 0 $. A similar behavior is present when proving the invariance of the effective four-dimensional Yang-Mills Lagrangian obtained after compactification of the fifth spatial extra dimension described  in Sec. \ref{YM5}. Finally, since Eqs.~\eqref{gtfmn2} are obtained from \eqref{gtsu2}, at the curvature level the SGTs and NSGTs are induced; in particular the SGTs of $F^{\bar{a}}_{\mu \nu}$, $Y_{\mu \nu}$, and $F^8_{\mu \nu}$ are
\begin{subequations}\label{sgtsu2f}
\begin{align}
\delta_{\!\!\trm{s}} F^{\bar{a}}_{\mu \nu} & = g\epsilon^{\bar{a}\bar{b}\bar{c}}F^{\bar{b}}_{\mu \nu}\alpha^{\bar{c}} \ ,\label{sgtfmn}\\
\delta_{\!\!\trm{s}}  Y_{\mu \nu} & = ig\frac{\sigma^{\bar{a}}}{2}Y_{\mu \nu}\alpha^{\bar{a}}\ , \label{sgtymn}\\
 \delta_{\!\!\trm{s}} F_{\mu\nu}^{8} & =0\ ,\label{sgtz}
\end{align}
\end{subequations}
which imply the previously enunciated fact: $F^{\bar{a}}_{\mu \nu}$, $Y_{\mu \nu}$, and $F^8_{\mu \nu}$ transform in the adjoint, fundamental, and trivial representation of $SU(2)$, respectively.

 It is interesting to note that one can fix the gauge for the $Y_\mu$ fields in a covariant way under the $SU(2)$ group. This is particularly useful in practical phenomenological applications~\cite{TM}. To do this, let
\begin{equation}
f^{\hat{a}}=\left(\delta^{\hat{a}\hat{b}}\partial_\mu-gf^{\hat{a}\hat{b}\bar{c}}A^{\bar{c}}_\mu\right)A^\mu_{\hat{b}}
\end{equation}
be the corresponding gauge-fixing functions. In the $SU(2)$ coordinates, these functions can be arranged in a doublet of this group as follows:
\begin{equation}
f_Y=D_\mu Y^\mu \, ,
\end{equation}
where $D_\mu$ is the covariant derivative in the fundamental representation of $SU(2)$.

We now proceed to study the Hamiltonian structure of the theory from the $SU(2)$ point of view. So, to describe the system in phase space terms, the following conjugate momenta are defined:
\begin{subequations}\label{pis}
\begin{align}
\piw^{\mu}_{\bar{a}}  & =  \dfrac{\partial \mc{L}_{SU(2)}}{\partial \dot{W}_{\mu}^{\bar{a}}}=F^{\mu 0}_{\bar{a}}\ ,\label{piw}\\
\piy^{\mu}_{r} & = \dfrac{\partial \mc{L}_{SU(2)}}{\partial \dot{Y}_{\mu}^{r}}=Y^{*\,\mu 0}_{r}\ ,\label{piy}\\
{\pi_{Y}^{*}}^{\mu\, r} & = \dfrac{\partial \mc{L}_{SU(2)}}{\partial \dot{Y}^{*}_{\mu\,r}}=Y^{\mu0\, r}\ ,\label{piystar}\\
\piz^{\mu} & = \dfrac{\partial \mc{L}_{SU(2)}}{\partial \dot{Z}_{\mu}}=F^{\mu0}_{8}\ .\label{piz}
\end{align}
\end{subequations}
It is important to notice that ${\pi_{W}}_{\bar{a}}^{\mu}$ are not the canonical momenta associated with the pure $SU(2)$ theory, whose Lagrangian is
\begin{equation}
\label{LSUP}
{\mc L}=-\frac{1}{4}W^{\bar{a}}_{\mu \nu}W^{\mu \nu}_{\bar{a}}\  ,
\end{equation}
and which leads to the conjugate momenta
\begin{equation}
p_{\bar{a}}^\mu=W^{\mu 0}_{\bar{a}} \ .
\end{equation}
These momenta differ from those derived from the Lagrangian in Eq.~(\ref{L2}), which are explicitly given by
\begin{equation}
\piw_{\bar{a}}^{\mu}=p_{\bar{a}}^{\mu}+ig\left(Y^{\dag\mu} \frac{\sigma^{\bar{a}}}{2}Y^0-Y^{\dag0} \frac{\sigma^{\bar{a}}}{2}Y^\mu\right)\ .
\end{equation}

The relations between the canonical momenta in the $ SU(3) $ and the $ SU(2) $ descriptions are
\begin{subequations}\label{ct}
\begin{align}
\pi_{\bar{a}}^\mu & = \piw_{\bar{a}}^\mu  \ , \label{pibara}\\
\pi_{4}^\mu & = \frac{1}{\sqrt{2}} \left(\piy^{\mu}_{1}+ \pi_{Y}^{*\;\mu 1} \right)\,  ,\ \pi_{5}^\mu = \frac{-i}{\sqrt{2}} \left(\piy^{\mu}_{1}- \pi_{Y}^{*\;\mu 1}\right) \, , \label{ct45}\\
\pi_{6}^\mu & = \frac{1}{\sqrt{2}} \left(\piy^{\mu}_{2}+ \pi_{Y}^{*\;\mu 2}\right)\,  ,\ \pi_{7}^\mu = \frac{-i}{\sqrt{2}} \left(\piy^{\mu}_{2}-\pi_{Y}^{*\;\mu 2}\right)\,  ,  \label{pi67}\\
\pi_{8}^\mu & = \pi_{Z}\  ,  \label{pi8}
\end{align}
\end{subequations}
whose inverses are
\begin{subequations}\label{invct}
\begin{align}
\piw^{\mu}_{\bar{a}} & =  \pi_{\bar{a}}^\mu\ , \label{ctW}\\
\piy^{\mu} & =(\piy^{\mu}_{1}\ \ \piy^{\mu}_{2})=\dfrac{1}{\sqrt{2}}(\pi^{\mu}_{4}+i\pi^{\mu}_{5}\ \  \pi^{\mu}_{6}+i\pi^{\mu}_{7})\ ,\label{ctYstar}\\
\pi_Y^{\dag\;\mu} & =
\begin{pmatrix}
\pi_{Y}^{*\;\mu 1}\\
\pi_{Y}^{*\;\mu 2}
\end{pmatrix} = \frac{1}{\sqrt{2}}
\begin{pmatrix}
\pi^{\mu}_{4}-i\pi^{\mu}_{5}\\
\pi^{\mu}_{6}-i\pi^{\mu}_{7}
\end{pmatrix}\ , \label{ctY}\\
\piz^{\mu} & =\pi^{\mu}_{8}\ . \label{ctZ}
\end{align}
\end{subequations}

From the conjugate momentum expressions Eq.~\eqref{pis} and using the notation Eq.~\eqref{invct}, one can readily recognize the primary constraints as
\begin{subequations}\label{pcsu2}
\begin{align}
\phi^{(1)}_{\bar{a}} &=\piw^{0}_{\bar{a}}\approx 0\ ,\label{pcbara}\\
\phi^{(1)}_Y&=\piy^{0} \approx 0\ ,\label{pcy}\\
\phi^{(1)\dag}_Y&=\pi_Y^{\dag\; 0} \approx 0\ ,\label{pcydag}\\
\phi^{(1)}_{Z} & = \piz^{0}\approx  0\ .\label{pcz}
\end{align}
\end{subequations}
Notice that $ \phi^{(1)}_Y $ and $  \phi^{(1)\dag}_Y$ are covariant and contravariant $SU(2)$ doublets, respectively. Since $ \piw^{\mu}_{\bar{a}} $ do not coincide with the conjugate momenta associated to the pure $ SU(2) $ theory Eq.~\eqref{LSUP}, the primary constraints $ \phi^{(1)}_{\bar{a}} $ differ from the primary constraints $ p_{\bar{a}}^{0} $ which emerge in the canonical analysis of Eq.~\eqref{LSUP}. The same observation will apply for the secondary constraints.

The primary Hamiltonian, which governs the evolution of the system, takes the form
\begin{equation}\label{phsu2}
\mc{H}^{(1)}_{SU(2)}={\mc H}_{SU(2)}+\mu^{\bar{a}} \phi^{(1)}_{\bar{a}}+\phi^{(1)}_{Y}\mu_{Y}+ \mu_{Y}^{\dag}\phi^{(1)\dag}_{Y}+\mu_{Z}\phi^{(1)}_{Z}\  .
\end{equation}
It corresponds to the sum of the canonical Hamiltonian
\begin{align}\label{eq:Hcsu2}
\mc{H}_{SU(2)}  =& \ \half \piw^{i}_{\bar{a}}\piw^{i}_{\bar{a}}+\piy^{i}\pi_{Y}^{\dag\;i}+\half\piz^{i}\piz^{i}+\quarter\left(F^{\bar{a}}_{ij}F^{ij}_{\bar{a}}+
2Y^{\dag}_{ij}Y^{ij}+F^{8}_{ij}F^{ij}_{8}\right)\nonumber\\
& -W^{\bar{a}}_{0}\phi_{\bar{a}}^{(2)}-\phi^{(2)\dag}_{Y}Y_{0}-Y_{0}^{\dag}\phi^{(2)}_{Y}-Z_{0}\phi^{(2)}_{Z}\ ,
\end{align}
and a linear combination of the primary constraints Eq.~\eqref{pcsu2} where the Lagrange multipliers $ \mu_{Y}  $, $ \mu_{Y}^\dag  $ and $ \mu_{Z} $ are
\begin{subequations}\label{lmsu2}
\begin{align}
\mu_{Y} & =\dfrac{1}{\sqrt{2}}
\begin{pmatrix}
{\mu}^{4}-i{\mu}^{5}\\
{\mu}^{6}-i{\mu}^{7}
\end{pmatrix}\ , \label{muY}\\
{\mu}_{Y}^{\dag} & =\dfrac{1}{\sqrt{2}}({\mu}^{4}+i{\mu}^{5}\ \  {\mu}^{6}+i{\mu}^{7})\ ,\label{muYstar}\\
{\mu}_{Z} & ={\mu}_{8}\ . \label{muZ}
\end{align}
\end{subequations}

By using the primary Hamiltonian  Eq.~\eqref{phsu2}, the consistency condition over the primary constraints Eq.~\eqref{pcsu2} yields the following secondary constraints:
\begin{subequations}\label{scsu2}
\begin{align}
\phi^{(2)}_{\bar{a}} &=\mc{D}^{\bar{a}\bar{b}}_{i}\piw^{i}_{\bar{b}}-ig\left(\piy^{i}\frac{\sgm^{\bar{a}}}{2}Y_{i}-Y_{i}^{\dag}\frac{\sgm^{\bar{a}}}{2}\pi_{Y}^{\dag\;i}
\right)\approx 0\ ,\label{scbara}\\
\phi^{(2)\dag}_Y&=\piy^{i}\left(\overset{\leftharpoonup}{D}_{i}+ig\frac{\sqrt{3}}{2}Z_{i}\right)-igY_{i}^{\dag}\left(\frac{\sgm^{\bar{a}}}{2}\piw^{i}_{\bar{a}}+\frac{\sqrt{3}}{2}\piz^{i}\right)\approx 0\ ,\label{scydag}\\
\phi^{(2)}_Y&=\left(D_{i}-ig\frac{\sqrt{3}}{2}Z_{i}\right)\pi^{\dag\; i}_{Y}+ig\left(\frac{\sgm^{\bar{a}}}{2}\piw^{i}_{\bar{a}}+\frac{\sqrt{3}}{2}\piz^{i}\right)Y_{i}\approx 0\ ,\label{scy}\\
\phi^{(2)}_{Z} & = \partial_{i}\piz^{i}-ig\frac{\sqrt{3}}{2}\left(\piy^{i}Y_{i}-Y_{i}^{\dag}\pi_{Y}^{\dag\; i}\right)\approx  0\ ,\label{scz}
\end{align}
\end{subequations}
where the action of $ \overset{\leftharpoonup}{D}_{\mu} $ on a contravariant $SU(2)$ doublet, say $ \piy^{\mu} $, is another  contravariant $SU(2)$ doublet defined by  $ \piy^{\mu}\overset{\leftharpoonup}{D}_{\mu}\equiv \partial_{\mu} \piy^{\mu}+ig\piy^{\mu}\frac{\sgm^{\bar{a}}}{2}W^{\bar{a}}_{\mu} $. The consistency condition applied to each secondary constraint yields no new constraints. It turns out, that all primary and secondary constraints do form a set of first-class constraints; in fact, the relevant Poisson brackets between these first-class constraints are
\begin{subequations}\label{gasu2}
\begin{align}
\{ \phi^{(2)}_{\bar{a}}[u],\phi^{(2)}_{\bar{b}}[v]\}_{SU(2)} & =g\eps_{\bar{a}\bar{b}\bar{c}}\,\phi^{(2)}_{\bar{c}}[uv]\ ,\label{ga-aa}\\
\{ \phi^{(2)}_{\bar{a}}[u],\phi^{(2)r}_{Y}[v]\}_{SU(2)}& =ig\frac{(\sgm^{\bar{a}})^{r}_{s}}{2}\,\phi^{(2)s}_Y[uv]\ , \label{ga-ay}\\
\{ \phi^{(2) r}_{Y}[u],\phi^{(2) s}_{Y}[v]\}_{SU(2)} &=g^{2}T^{rs}_{pq}\int d^{3}x \,(uv)(\mathbf{x})\big(\pi_{Y}^{*\;iq}Y_{i}^{p}-\pi_{Y}^{*\;ip}Y_{i}^{q}\big)(\mathbf{x})\ , \label{ga-yy}\\
\{ \phi^{(2) r}_{Y}[u],\phi^{(2)*}_{Ys}[v]\}_{SU(2)} &=ig\left(\frac{(\sgm^{\bar{a}})^{r}_{s}}{2}\,\phi^{(2)}_{\bar{a}}[uv]+\frac{\sqrt{3}}{2}\dlt^{r}_{s}\phi^{(2)}_{Z}[uv]\right)\nonumber\\
& \ \ +g^{2}(T^{rs}_{pq}-T^{rs}_{qp})\int d^{3}x \,(uv)(\mathbf{x})\big(\piy_{s}^{i}Y_{i}^{p}-{Y}^{*}_{is}\pi_{Y}^{*\;ip}\big)(\mathbf{x})\ , \label{ga-yystar}\\
\{ \phi^{(2)r}_{Y}[u],\phi^{(2)}_{Z}[v]\}_{SU(2)} &=-\dfrac{ig\sqrt{3}}{2}\phi^{(2) r}_Y[uv] \ , \label{ga-yz}
\end{align}
\end{subequations}
where $ \{\cdot,\cdot\}_{SU(2)} $ denotes the Poisson bracket that involves the $ SU(2) $ phase space coordinates. Due to the symmetries present in the lower indices of $ T^{rs}_{pq} $, one finds that the terms proportional to $ g^{2} $ on the right hand side of Eq.~\eqref{ga-yy} and Eq.~\eqref{ga-yystar}  do not contribute to the occurrence of tertiary constraints; instead these terms identically vanish,  and the Poisson brackets among all the constraints give a linear combination of constraints themselves. A more elegant argument to show that such terms must identically vanish  on the whole phase space is the following: Notice that Eq.~\eqref{pt1} and Eq.~\eqref{ct} define a canonical transformation in the ordinary sense \cite{golbk}, and hence $ \{\cdot,\cdot\}_{SU(3)} = \{\cdot,\cdot\}_{SU(2)} $. Moreover, it is easy to see that this canonical transformation maps the primary constraints Eq.~\eqref{pc3} onto Eq.~\eqref{pcsu2}; hence the primary Hamiltonian in the $ SU(3) $ phase space coordinates Eq.~\eqref{phsu3} becomes the corresponding Hamiltonian in the $ SU(2) $ coordinates Eq.~\eqref{phsu2}. As a consequence, the set  of secondary constraints in both formalisms must match under the canonical transformation. Indeed, this can be proved by direct calculation. Since exclusively the primary Hamiltonian is employed to evolve the constraints in time through the Poisson bracket, one concludes that the Dirac algorithm in the $ SU(2) $ formulation must lack of tertiary constraints just as it does in the $ SU(3) $ formulation; this fact rules out the presence of the extra-terms proportional to $ g^{2} $ in the gauge algebra Eq.~\eqref{gasu2}. In conclusion the canonical transformation  defined by Eq.~\eqref{pt1} and Eq.~\eqref{ct} maps each stage of the Dirac algorithm in the $ SU(3) $ formulation into the corresponding stage in the $ SU(2) $ one. Notice that the number of physical degrees of freedom of the $ SU(2) $ effective theory matches with the corresponding number of the pure $ SU(3) $ Yang-Mills theory.

We end the Hamiltonian analysis from the $ SU(2) $ perspective by calculating the gauge generator $ G $ \cite{cas82}. This generator is linear in all first-class constraints Eq.~\eqref{pcsu2} and Eq.~\eqref{scsu2} with coefficients of the primary ones related to that of the secondary ones; the relation among the coefficients is obtained by imposing the condition that the total time derivative of $ G $,
\begin{equation*}
\frac{\partial G}{\partial t} +\{G,\mc{H}_{SU(2)}\}_{SU(2)}\ ,
\end{equation*}
must be a linear combination of the primary constraints only \cite{pon05}. As a consequence one gets
\begin{align}\label{ggsu2}
G = & \ \big[\mc{D}_{0}^{\bar{a}\bar{b}}\al^{\bar{b}}-ig\big(\bt^{\dag}\dfrac{\sgm^{\bar{a}}}{2}Y_{0}-Y_{0}^{\dag}\dfrac{\sgm^{\bar{a}}}{2}\bt\big)\big]\phi^{(1)}_{\bar{a}}+\phi^{(1)}_{Y}\big[\big(D_{0}-ig\frac{\sqrt{3}}{2}Z^{0}\big)\bt+ig\big(\dfrac{\sgm^{\bar{a}}}{2}\al^{\bar{a}}-\frac{\sqrt{3}}{2}\al_{Z}\big)Y_{0}\big]\nonumber\\
&\ +\big[\bt^{\dag}\big(\overset{\leftharpoonup}{D}_{0}+ig\frac{\sqrt{3}}{2}Z^{0}\big)-igY_{0}^{\dag}\big(\dfrac{\sgm^{\bar{a}}}{2}\al^{\bar{a}}-\frac{\sqrt{3}}{2}\al_{Z}\big)\big]\phi^{(1)\dag}_{Y}+\big[\partial_{0}\al_{Z}+ig\big(\bt Y_{0}^{\dag}-\bt^{\dag}Y_{0}\big)\big]\phi^{(1)}_{Z}\nonumber\\
&\ -\al^{\bar{a}}\phi^{(2)}_{\bar{a}}-\bt^{\dag}\phi^{(2)}_{Y}-\phi^{(2)\dag}_{Y}\bt-\al_{Z}\phi^{(2)}_{Z}\ .
\end{align}
This gauge generator is the sum of $ G_{\!\!\trm{s}}\equiv G\vert_{\bt=0,\al_{Z}=0} $ and  $ G_{\!\!\trm{ns}}\equiv G\vert_{\al^{\bar{a}}=0} $ which independently generate the SGTs and NSGTs, Eqs.~\eqref{sgt2} and \eqref{nsgtsu2}, respectively, via the Poisson brackets\begin{subequations}\label{gtcassu2}
\begin{align}
\dlt_{\!\!\trm{s}} W^{\bar{a}}_{\mu} & = \{ W^{\bar{a}}_{\mu},G_{\!\!\trm{s}}\}_{SU(2)},\ \dlt_{\!\!\trm{s}} Y_{\mu}  = \{ Y_{\mu},G_{\!\!\trm{s}}\}_{SU(2)}, \ \dlt_{\!\!\trm{s}} Z_{\mu} = \{ Z_{\mu},G_{\!\!\trm{s}}\}_{SU(2)}\ ,\label{sgtwyz}\\
\dlt_{\!\!\trm{ns}} W^{\bar{a}}_{\mu} & = \{ W^{\bar{a}}_{\mu},G_{\!\!\trm{ns}}\}_{SU(2)},\ \dlt_{\!\!\trm{ns}} Y_{\mu}  = \{ Y_{\mu},G_{\!\!\trm{ns}}\}_{SU(2)}, \ \dlt_{\!\!\trm{ns}} Z_{\mu} = \{ Z_{\mu},G_{\!\!\trm{ns}}\}_{SU(2)} \label{nsgtwyz}\ .
\end{align}
\end{subequations}
From these transformation laws and the constraint algebra Eq.~\eqref{gasu2}, it is straightforward to see that on the constraint surface the Lie algebra among SGTs and NSGTs can be  summarized as follows:
\begin{equation}\label{SGT-NSGT alg}
[\!\trm{SGT},\!\!\trm{SGT}\,]=\!\!\trm{SGT},\quad[\!\trm{SGT},\!\!\trm{NSGT}\,]=\!\!\trm{SGT}+\!\!\trm{NSGT},\quad[\!\trm{NSGT},\!\!\trm{NSGT}\,]=\!\!\trm{SGT}+\!\!\trm{NSGT},
\end{equation}
where $ [\cdot,\cdot] $ denotes a Lie product. The first of these relations closes with structure constants and specifically follows from the Lie subalgebra Eq.~\eqref{ga-aa}, therefore exponentiation of SGTs provides a Lie group which in fact corresponds to $ SU(2) $. Since the Lie product of  NSGTs does not close, they do not exponentiate into a group. The complete transformations \eqref{gtsu2} are duly reproduced by the addition $ \dlt=\dlt_{\!\!\trm{s}}+\dlt_{\!\!\trm{ns}}\, $. It is worth noticing that the gauge generator Eq.~\eqref{ggsu2} is the image of the gauge generator Eq.~\eqref{gg3} under the canonical transformation defined by  Eqs.~\eqref{pt1} and \eqref{ct}.

To conclude this subsection, we would like to emphasize the following: A hidden symmetry arises when an admissible canonical transformation is introduced. The canonical transformation is admissible in the sense that it maps well defined objects under some group $G$ to well defined  objects of a subgroup $H$ of $ G $. The gauge symmetry, which is manifest in $G$, is hidden in $H$. The gauge symmetries with respect to the group $G$ that appear hidden from the $H$ perspective are those associated with the generators of $G$ that do not generate $H$. This is true independently of whether or not the $G$ group is spontaneously broken down into $H$. In our toy model $ G=SU(3) $ and $ H=SU(2) $; after the canonical transformation, only the fields  $ W^{\bar{a}} _{\mu}=A_{\mu}^{\bar{a}}$ explicitly continue being gauge fields under $ H $. The rest of the fields, $ Y_{\mu} $, $ Y_{\mu}^{\dag} $ and $ Z_{\mu} $, fulfill very different transformation laws under $ H $; nevertheless, the latter fields can be mapped back with the canonical transformation to gauge fields with respect to $ G $. This result is crucial for our study of passing from the $SU(N,{\mc M}^{5})$ gauge group description to the  $SU(N,{\mc M}^4)$ one via compactification, as in this case the phenomenon of spontaneous symmetry breaking is not present. Note that in this subtler case $SU(N,{\mc M}^4)$ is a subgroup of $SU(N,{\mc M}^{5})$ not due to a difference in the number of generators, which is the same indeed, but because the gauge parameters of the group $SU(N,{\mc M}^{5})$ are restricted to take values on the submanifold $ \mc{M}^{4} $ of $ \mc{M}^{5} $. We will show that there exists an admissible canonical transformation in this case.

\section{The {$SU(3)$} Yang-Mills theory with spontaneous symmetry breaking}
\label{SU3SSB}
We now proceed to extend the study of the previous section to the case when the $SU(3)$ group is spontaneously broken into the $SU(2)$ in the usual sense. One of the two main purposes is to contrast the notion of hidden symmetry induced by a suitable canonical transformation with that coming from SSB. The other is to show how a specific NSGT can be used to define the unitary gauge.  In this scenario, we will be able to make a precise analogy of this procedure with a similar one used in the context of extra dimensions.

\subsection{The {$SU(3)$} perspective of the model}
To carry out the mentioned SSB, we add to the pure $SU(3)$ theory given by the Lagrangian in Eq.~(\ref{L3}) a renormalizable scalar sector $ \mc{L}_{\Phi} $, so that
\begin{equation}
\label{LS3}
{\mc L}_{SU(3),\Phi}={\mc L}_{SU(3)}+{\mc L}_\Phi \, ,
\end{equation}
where
\begin{equation}
{\mc L}_\Phi=(D_\mu \Phi)^\dag (D^\mu \Phi)-V(\Phi^\dag,\Phi)\, .
\end{equation}
 In this expression $D_\mu=\partial_\mu -ig\frac{\lambda^a}{2}A^a_\mu$ is the covariant derivative in the fundamental representation of $SU(3)$,\footnote{We trust that no confusion will arise with the symbol $ D_{\mu} $ already used for the covariant derivative of $ SU(2) $ in its fundamental representation, as we think one can infer the nature of the covariant derivative depending on which object this is acting on.} and $\Phi$ is a complex contravariant Poincar\'{e} scalar triplet of $SU(3)$. In addition, $V(\Phi^\dag,\Phi)$ is the renormalizable scalar potential given by
\begin{equation}
V(\Phi^\dag,\Phi)=\mu^2\left(\Phi^\dag \Phi\right)+\lambda\left(\Phi^\dag \Phi\right)^2 \, .
\end{equation}
It is straightforward to show that the Lagrangian in Eq.~\eqref{LS3} is simultaneously invariant under Eq.~\eqref{gt} and the infinitesimal rotation of the triplet $ \Phi $ in the isospin space,
\begin{equation}\label{gtfi3}
    \delta \Phi=-i\alpha^{a}\left(\frac{\lambda^a}{2}\Phi\right)\, .
\end{equation}

 The gauge symmetries of the Lagrangian in Eq.~\eqref{LS3} will be reflected in the occurrence of first-class constraints in the Hamiltonian setting. In order to formulate the theory in phase space terms, in addition to the canonical pairs $ (A_{\mu}^{a},\pi^{\mu}_{a}) $, \cf~Eqs.~\eqref{pis3}, the conjugate pairs $ (\Phi,\pi) $ and $ (\Phi^{\dag},\pi^{\dag}) $ must be introduced, where
 \begin{subequations}\label{pisfi3}
\begin{align}
\pi & =\dfrac{\partial \mc{L}_{\Phi}}{\partial \dot{\Phi}}=(D_0\Phi)^\dag \ , \label{pi}\\
\pi^{\dag} & =\dfrac{\partial \mc{L}_{\Phi}}{\partial \dot{\Phi}^{\dag}}=D_0\Phi\ .\label{pistar}
\end{align}
\end{subequations}
Note that $ \pi $ and $ \pi^{\dag} $ correspond to covariant and contravariant $ SU(3) $ triplets, respectively. From the Eqs.~\eqref{pisfi3} the velocities $ \dot{\Phi}^{\dag} $ and $ \dot{\Phi} $ are expressible in terms of phase space variables; therefore they do not give rise to more primary constraints in addition to those defined in \eqref{pc3}. To bring uniformity  into the present section, primary constraints will be denoted by $ \vfi_{a}^{(1)}\equiv\phi_{a}^{(1)} $. The incorporation of the scalar sector into the pure $ SU(3) $ Yang-Mills Lagrangian does not have influence upon the primary constraints of the pure theory alone.

The canonical Hamiltonian associated with Eq.~\eqref{LS3} will be the sum of Eq.~\eqref{ch3} and the contribution from the Higgs sector $ \mc{L}_{\Phi} $, namely
\begin{equation}\label{ch3ssb}
\mc{H}_{SU(3),\Phi}=\mc{H}_{SU(3)}+\mc{H}_\Phi\ ,
\end{equation}
where
\begin{equation}\label{chhiggs}
\mc{H}_{\Phi} = \ \pi\pi^{\dag}+ig A^{a}_{0}\big(\pi\frac{\lm^{a}}{2}\Phi-\Phi^{\dag}\frac{\lm^{a}}{2}\pi^{\dag}\big)-\big(D_{i}\Phi\big)^{\dag}(D^{i}\Phi)+V(\Phi,\Phi^{\dag})\ .
\end{equation}
Notice that the term linear in $ A^{a}_{0} $ will modify the secondary constraints that are produced in the absence of the Higgs sector. Indeed, the primary Hamiltonian
\begin{equation}\label{phsu3ssb}
\mc{H}^{(1)}_{SU(3),\Phi}=\mc{H}_{SU(3), \Phi}+\mu^{a}\vfi_{a}^{(1)}
\end{equation}
allows us to obtain the consistency condition on the primary constraints Eq.~\eqref{pc3} providing the following secondary constraints:
\begin{equation}\label{sc3ssb}
\varphi^{(2)}_{a}\equiv \phi^{(2)}_{a}-ig\big(\pi\frac{\lm^{a}}{2}\Phi-\Phi^{\dag}\frac{\lm^{a}}{2}\pi^{\dag}\big)\approx 0 \ ,
\end{equation}
where $ \phi^{(2)}_{a} $ corresponds to the secondary constraints Eq.~\eqref{sc3} conveyed by the pure $ SU(3) $ Yang-Mills theory. The consistency requirement on  $ \varphi^{(2)}_{a} $ does not bring more constraints, ending with the Dirac algorithm. The primary and secondary constraints of the theory, Eqs.~\eqref{pc3} and  \eqref{sc3ssb}, form a set of first-class constraints; the nonvanishing Poisson brackets between the constraints reveal the $ SU(3) $ symmetry of the theory
\begin{equation}\label{gasu3ssb}
\{\varphi^{(2)}_a[u] , \varphi^{(2)}_b [v]\}_{SU(3)}=gf_{abc}\varphi^{(2)}_c[uv] \ ,
\end{equation}
where $ \{\cdot,\cdot\}_{SU(3)} $ is the Poisson bracket in the $ SU(3) $ formulation which takes into account the conjugate pairs $ (A_{\mu}^{a},\pi_{a}^{\mu}) $, $ (\Phi,\pi) $ and $ (\Phi^{\dag},\pi^{\dag}) $.
Since only secondary constraints are modified by the Higgs sector, one expects that once the SSB  of $ SU(3) $ into $ SU(2) $ operates, the affected constraints will only be  the secondary ones.

Before going into the $ SU(2) $ formulation of the theory, the gauge generator is presented. Linear in all first-class constraints, this corresponds to
\begin{equation}\label{gg3ssb}
G= (\mc{D}^{ab}_{0}\al^{b})\varphi^{(1)}_{a}-\al^{a}\varphi^{(2)}_{a}\ .
\end{equation}
Notice that the scalar contribution in the secondary constraints Eq.~\eqref{sc3ssb} is responsible for the appropriate transformation law that the scalar fields must follow, \cf~Eq.~\eqref{gtfi3}; in fact,
\begin{subequations}\label{gtcas3ssb}
\begin{align}
\dlt A_{\mu}^{a} & = \{A^{a}_{\mu},G\}_{SU(3)} \label{gtcasa3ssb}\\
\dlt \Phi & =\{\Phi,G\}_{SU(3)} \label{gtcasfi3ssb}
\end{align}
\end{subequations}
faithfully reproduce Eqs.~\eqref{gt} and \eqref{gtfi3} --that is, the symmetries of the theory.

\subsection{SSB from the {$SU(3)$}  perspective}
\label{SSBSU3}
In this subsection we revisit the SSB~\cite{GT} from what we have referred to as the  $ SU(3) $ perspective. We consider the case $\mu^2<0$, in which the vacuum is infinitely degenerate, so the theory presents SSB.

The extremum at $ \Phi=0 $ is not considered. We may presume that the expectation value of $ \Phi $ in the vacuum does not vanish. The energy of the system is minimal on all the points of the spherical surface given by
\begin{equation}\label{minssb}
{\Phi}^{\dag}_{\!\!\trm{min}} \Phi_{\!\!\trm{min}}=-\frac{\mu^{2}}{2\lm}\equiv v^2\ .
\end{equation}
All points on these surface are physically equivalent because they are connected through $SU(3)$ transformations. To break down $SU(3)$ into $SU(2)$, one chooses a particular direction $\Phi_{\!\!\trm{min}}$ such that
\begin{subequations}\label{su3-to-su2}
\begin{align}
\frac{\lambda^{\bar{a}}}{2}\Phi_{\!\!\trm{min}} &=0 \  , \\
\frac{\lambda^{\hat{a}}}{2}\Phi_{\!\!\trm{min}} &\neq 0 \  , \\
\frac{\lambda^8}{2} \Phi_{\!\!\trm{min}} &\neq 0 \  .
\end{align}
\end{subequations}
The isotropy group, the one corresponding to unbroken symmetries, at $ \Phi_{\!\!\trm{min}} $ is $ SU(2) $. It is convenient to choose a representative of the solutions to the Eq.~\eqref{minssb} as $ \Phi_{\!\!\trm{min}}^{\dag} = (0\ \ 0 \ \ v)$. This choice means that five generators of $SU(3)$, namely, $\frac{\lambda^{\hat{a}}}{2}$ and $\frac{\lambda^{8}}{2}$, are broken.

Within this formulation, two cases clearly arise depending on the nature of the gauge parameters $ \al^{a} $ (\cf~Eq.~\eqref{gt}). These are
\begin{itemize}
\item[$ (i) $] \emph{The Goldstone Theorem}~\cite{GT}. Assuming the parameters $ \al^{a} $ to be constant functions on Minkowski space, the invariant Lagrangian corresponds to
\begin{equation*}
\mc{L}_{SU(3),H}=(\partial_{\mu} \Phi)^{\dag}(\partial^{\mu}\Phi)-V(\Phi^{\dag},\Phi)\ .
\end{equation*}
When the theory is subjected to the translation $\Phi\mapsto \varphi\equiv\Phi-\Phi_{\!\!\trm{min}}\, $, there arise five real massless scalars. These correspond to $ \varphi^{1} $, $ \varphi^{2} $ and the imaginary part of $ \varphi^{3} $ denoted as $ \phi_{Z} $. In addition, a massive scalar $ H$ emerges, identified as the real part of $ \varphi^{3} $, that quantifies the normal excitations to the surface of the minimal energy. Hence, associated with each broken generator of $ SU(3) $ there is a massless scalar or Goldstone boson.

\item[$ (ii)$] \emph{The Higgs Mechanism}~\cite{HM}.  Assuming the parameters $ \al^{a} $ to be  nonconstant functions on Minkowski space, the invariant Lagrangian corresponds to Eq.~\eqref{L3}. In this case, besides the presence of five pseudo-Goldstone bosons, five massive gauge bosons ($ A^{\hat{a}}_{\mu} $ and $ A^{8}_{\mu} $) arise. This is the celebrated Higgs mechanism. In this scenario, the pseudo-Goldstone bosons represent spurious degrees of freedom, as they can be removed from the theory in a special gauge, known as unitary gauge. In the following section, we will show that this mechanism has a natural description in the $ SU(2) $ coordinates, and that the unitary gauge can be understood as the action of fixing the parameters within what will be defined as NSGT on the scalar fields, Eq.~\eqref{nsgtfi}.
\end{itemize}

\subsection{The {$SU(2)$}  perspective of the model}

In this subsection the description of the field theory Eq.~\eqref{LS3} from the $ SU(2) $ perspective is achieved. The pure $ SU(3) $ Yang-Mills sector $ \mc{L}_{SU(3)} $ is mapped, by means of the point transformation Eq.~\eqref{pt1}, into $ \mc{L}_{SU(2)} $ Eq.~\eqref{L2}, and the scalar sector $ \mc{L}_{\Phi} $ is mapped onto $ \mc{L}_{\phi} $ by decomposing the $ SU(3) $ triplet $ \Phi $ into an $ SU(2) $ doublet and a scalar,
\begin{subequations}\label{ptfi}
\begin{align}
\begin{pmatrix}
 \Phi^{1} \\
\Phi^{2}
\end{pmatrix}
 &  =
\begin{pmatrix}
 \phi^1  \\
\phi^2
\end{pmatrix}  ,\label{ptfis}\\
\Phi^3 & = \phi^0\ .\label{ptfi0}
\end{align}
\end{subequations}
Therefore, the Lagrangian in Eq.~\eqref{LS3} is recast in terms of well defined objects under the action of $ SU(2) $,
\begin{equation}\label{L2SSB}
\mc{L}_{SU(2),\phi}=\mc{L}_{SU(2)}+\mc{L}_{\phi},
\end{equation}
where the Higgs sector becomes
\begin{equation}\label{Lfi}
\mc{L}_{\phi}=(D_\mu \Phi)^\dag (D^\mu \Phi)\big\vert_{\substack{
\Phi \,\to\, \phi \hfill\\
 A_{\mu}\to \,W_{\mu},Y_{\mu},Y_{\mu}^{\dag}, Z_{\mu} }}
 +V(\Phi,\Phi^{\dag})\big\vert_{\Phi\, \to\, \phi}\ .
\end{equation}

Gauge invariances of the theory in this formulation correspond to Eq.~\eqref{LS3}  together with
\begin{subequations}\label{gtfi2}
\begin{align}
\dlt\phi & =-i\left(\frac{\sgm^{\bar{a}}}{2}\al^{\bar{a}}+\frac{1}{2\sqrt{3}}\al_{Z}\right)\phi-\dfrac{i}{\sqrt{2}}\phi^{0}\bt\ , \label{gtfiy2}\\
\dlt\phi^{0} & = - \dfrac{i}{\sqrt{2}}\bt^{\dag}\phi+\frac{i}{\sqrt{3}}\al_{Z}\phi^{0}\ .\label{gtfi02}
\end{align}
\end{subequations}
Notice that in the scalar sector of the theory, the SGTs and NSGTs also naturally arise. Indeed
\begin{subequations}\label{eq:sgt-nsgt-fi}
\begin{align}
\dlt_{\!\!\trm{s}}\phi &= -i\frac{\sgm^{\bar{a}}}{2}\al^{\bar{a}}\phi \, ,\  \dlt_{\!\!\trm{s}}\phi^{0}=0\ ;\label{sgtfi}\\
\dlt_{\!\!\trm{ns}}\phi &= -\frac{i}{2}\left(\frac{1}{\sqrt{3}}\al_{Z}\phi+{\sqrt{2}}\phi^{0}\bt\right) \, ,\  \dlt_{\!\!\trm{ns}}\phi^{0}= - \dfrac{i}{\sqrt{2}}\bt^{\dag}\phi+\frac{i}{\sqrt{3}}\al_{Z}\phi^{0}\ .\  \label{nsgtfi}
\end{align}
\end{subequations}

We now proceed to the Hamiltonian formulation associated to the singular Lagrangian Eq.~\eqref{L2SSB}. Since the scalar sector does not contain spacetime derivatives  of  either gauge fields $ W^{\bar{a}}_{\mu} $, or $ SU(2) $ doublets $ Y_{\mu} $, or the scalar $ Z_{\mu} $, the canonical conjugate momentum associated with each  of these fields coincides with those defined in Sec.~\ref{SU2}. Hence, the conjugate momenta in the $ SU(2) $ formulation are given by Eqs.~\eqref{pis} and
\begin{subequations}\label{pisfi2}
\begin{align}
\pi_{\phi} & =\dfrac{\partial \mc{L}_{\phi}}{\partial \dot{\phi}}=\phi^{\dag}\left(\overset{\leftharpoonup}{D}_{0}+\frac{ig}{2\sqrt{3}}Z_{0}\right)+\dfrac{ig}{\sqrt{2}}\phi^{0\, *}\,Y_{0}^{\dag}\  ,\label{pifi}\\
\pi_{0} & =\dfrac{\partial \mc{L}_{\phi}}{\partial \dot{\phi}^{0}}=\left(\partial_{0} -\dfrac{ig}{\sqrt{3}}Z_{0}\right)\phi^{0\, *} +\dfrac{ig}{\sqrt{2}}\phi^{\dag}\,Y_{0}\ , \label{pi0}\\
\pi_{\phi}^{\dag} & =\dfrac{\partial \mc{L}_{\phi}}{\partial \dot{\phi}^{\dag}}=\left({D}_{0}-\frac{ig}{2\sqrt{3}}Z_{0}\right)\phi-\dfrac{ig}{\sqrt{2}}\phi^{0}\,Y_{0}\ ,\label{pifistar}\\
 \pi_{0}^{*} & =\dfrac{\partial \mc{L}_{\phi}}{\partial \dot{\phi}^{0\, *}}=\left(\partial_{0} +\dfrac{ig}{\sqrt{3}}Z_{0}\right)\phi^{0} -\dfrac{ig}{\sqrt{2}}Y_{0}^{\dag}\,\phi\ . \label{pifipi0star}
\end{align}
\end{subequations}
It is worth noticing that $ \pi_{\phi} $ and $ \pi_{\phi}^{\dag} $ are covariant and contravariant $ SU(2) $ doublets, respectively, whereas, $ \pi_{0} $ and its complex conjugate are $ SU(2) $ scalars. The relations among conjugate momenta \eqref{pisfi2} and the corresponding objects Eq.~\eqref{pisfi3} are
\begin{subequations}\label{ctpifi}
\begin{align}
\pi_{\phi} & =(\pi_{\phi}^{1}\ \ \pi_{\phi}^{2})=(\pi^{1}\ \ \pi^{2})\ , \label{ctpisu2}\\
\pi_{0} & = \pi^{3} \ . \label{ctpi3}
\end{align}
\end{subequations}

As expected, the scalar sector of the theory does not bring additional constraints into the $ SU(2) $ formalism either.  Instead of going through the Dirac formalism using the Poisson bracket $ \{\cdot,\cdot\}_{SU(2)}$, that in this case would include  also the canonical pairs $ (\phi,\pi_{\phi}) $, $ (\phi^{0},\pi_{0}) $, $ (\phi^{\dag},\pi^{\dag}_{\phi}) $, and $ (\phi^{0 *},\pi_{0}^{*}) $, we will make use of the arguments given after Eqs.~\eqref{gtcassu2} in the following way: First, notice that Eqs.~\eqref{pt1}, \eqref{ct}, \eqref{ptfi} and \eqref{ctpifi} define a canonical transformation from $ SU(3) $ to $ SU(2) $ coordinates; therefore $ \{\cdot,\cdot\}_{SU(3)}= \{\cdot,\cdot\}_{SU(2)}$. Second, the canonical transformation maps the set of primary constraints $\{ \vfi^{(1)}_{a} \}$ onto the set of primary constraints $ \{\vfi^{(1)}_{\bar{a}}\equiv\phi^{(1)}_{\bar{a}},\vfi^{(1)}_{Y}\equiv\phi^{(1)}_{Y},\vfi^{(1)}_{Z}\equiv\phi^{(1)}_{Z} \} $; the transformation hence recasts the primary  Hamiltonian Eq.~\eqref{phsu3ssb} in terms of $ SU(2) $ variables as follows:
\begin{equation}\label{ph2ssb}
\mc{H}^{(1)}_{SU(2),\phi}=\mc{H}_{SU(2)}+\mc{H}_{\phi}+\mu^{\bar{a}} \vfi^{(1)}_{\bar{a}}+\vfi^{(1)}_{Y}\mu_{Y}+ \mu_{Y}^{\dag}\vfi^{(1)\dag}_{Y}+\mu_{Z}\vfi^{(1)}_{Z}\
\end{equation}
where $ \mc{H}_{SU(2)} $ is given by Eq.~\eqref{L2} and $ \mc{H}_{\phi} $ is the Legendre transformation of $ \mc{L}_{\phi} $. As a consequence of these two observations, the set of secondary constraints that emerges in the $ SU(3) $ viewpoint must be faithfully mapped onto the set  of secondary constraints given in terms of the $ SU(2) $ coordinates. These are
\begin{subequations}\label{sc2ssb}	
\begin{align}
\vfi^{(2)}_{\bar{a}} & = \phi^{(2)}_{\bar{a}}-ig \left(\pi_{\phi}\frac{\sgm^{\bar{a}}}{2}\phi-\phi^{\dag}\frac{\sgm^{\bar{a}}}{2}\pi_{\phi}^{\dag}\right)\approx 0\ ,\label{sc2ssbbara}\\
\vfi^{(2)\dag}_{Y} & = \phi^{(2)\dag}_{Y}+\frac{ig}{\sqrt{2}}\left(\pi_{0}^{*}\phi^{\dag}-\phi^{0}\pi_{\phi}\right)\approx 0\ , \label{sc2ssbbfiystar}\\
\vfi^{(2)}_{Y} & = \phi^{(2)}_{Y}-\frac{ig}{\sqrt{2}}\left(\pi_{0}\phi-\phi^{0\, *}\pi_{\phi}^{\dag}\right)\approx 0\ , \label{sc2ssbfiy}\\
\vfi^{(2)}_{Z} & = \phi^{(2)}_{Z}-\frac{ig}{\sqrt{3}}\left(\phi^{0\, *}\pi_{0}^{*}-\pi_{0}\phi^{0}+\frac{1}{2}(\pi_{\phi}\phi-\phi^{\dag}\pi_{\phi}^{\dag})\right)\approx 0\ ,\label{sc2ssbz}
\end{align}
\end{subequations}
where $ \phi^{(2)}_{\bar{a}} $, $ \phi^{(2)\dag}_{Y} $, $ \phi^{(2)}_{Y} $ and $ \phi^{(2)}_{Z} $ are given by Eqs.~\eqref{scsu2}. Indeed, this can be proved by direct calculation. Finally, the set of equations that define the gauge algebra Eq.~\eqref{gasu3ssb} can be expressed in terms of $ SU(2) $ variables using only the canonical transformation. The nonvanishing Poisson brackets are
\begin{subequations}\label{gasu2ssb}
\begin{align}
\{ \varphi^{(2)}_{\bar{a}}[u],\varphi^{(2)}_{\bar{b}}[v]\}_{SU(2)} & =g\eps_{\bar{a}\bar{b}\bar{c}}\,\varphi^{(2)}_{\bar{c}}[uv]\ ,\label{gassb-aa}\\
\{ \varphi^{(2)}_{\bar{a}}[u],\varphi^{(2)r}_{Y}[v]\}_{SU(2)}& =ig\frac{(\sgm^{\bar{a}})^{r}_{s}}{2}\,\varphi^{(2)s}_Y[uv]\ , \label{gassb-ay}\\
\{ \varphi^{(2) r}_{Y}[u],\varphi^{(2)*}_{Ys}[v]\}_{SU(2)} &=ig\left(\frac{(\sgm^{\bar{a}})^{r}_{s}}{2}\,\varphi^{(2)}_{\bar{a}}[uv]+\frac{\sqrt{3}}{2}\dlt^{r}_{s}\varphi^{(2)}_{Z}[uv]\right)\ ,\label{gassb-yystar}\\
\{ \varphi^{(2)r}_{Y}[u],\varphi^{(2)}_{Z}[v]\}_{SU(2)} &=-\dfrac{ig\sqrt{3}}{2}\varphi^{(2) r}_Y[uv] \ . \label{gassb-yz}
\end{align}
\end{subequations}

Since the canonical transformation connects the Dirac algorithm unfolded in the two different sets of coordinates at each step, we have that the gauge generator Eq.~\eqref{gg3ssb} must be translated into the corresponding one in the $ SU(2) $ variables, namely
\begin{align}\label{ggsu2ssb}
G = & \ \big[\mc{D}_{0}^{\bar{a}\bar{b}}\al^{\bar{b}}-ig\big(\bt^{\dag}\dfrac{\sgm^{\bar{a}}}{2}Y_{0}-Y_{0}^{\dag}\dfrac{\sgm^{\bar{a}}}{2}\bt\big)\big]\varphi^{(1)}_{\bar{a}}+\varphi^{(1)}_{Y}\big[\big(D_{0}-ig\frac{\sqrt{3}}{2}Z^{0}\big)\bt+ig\big(\dfrac{\sgm^{\bar{a}}}{2}\al^{\bar{a}}-\frac{\sqrt{3}}{2}\al_{Z}\big)Y_{0}\big]\nonumber\\
&\ +\big[\bt^{\dag}\big(\overset{\leftharpoonup}{D}_{0}+ig\frac{\sqrt{3}}{2}Z^{0}\big)-igY_{0}^{\dag}\big(\dfrac{\sgm^{\bar{a}}}{2}\al^{\bar{a}}-\frac{\sqrt{3}}{2}\al_{Z}\big)\big]\varphi^{(1)\dag}_{Y}+\big[\partial_{0}\al_{Z}+ig\big(\bt Y_{0}^{\dag}-\bt^{\dag}Y_{0}\big)\big]\varphi^{(1)}_{Z}\nonumber\\
&\ -\al^{\bar{a}}\varphi^{(2)}_{\bar{a}}-\bt^{\dag}\varphi^{(2)}_{Y}-\varphi^{(2)\dag}_{Y}\bt-\al_{Z}\varphi^{(2)}_{Z}\ ,
\end{align}
from which the sectors that independently generate SGTs, $ G_{\!\!\trm{s}}\equiv G\vert_{\bt=0,\al_{Z}=0}$, and NSGT, $  G_{\!\!\trm{ns}}\equiv G\vert_{\al^{\bar{a}}=0}$, are easily identified. Notice that it is due to the terms depending on the Higgs sector in each secondary constraint that Eqs.~\eqref{eq:sgt-nsgt-fi} are suitably recovered from the following brackets:
\begin{subequations}\label{gtcassu2ssb}
\begin{align}
\dlt_{\!\!\trm{s}} \phi & = \{ \phi,G_{\!\!\trm{s}}\}_{SU(2)},\ \dlt_{\!\!\trm{s}} \phi^{0}  = \{ \phi^{0},G_{\!\!\trm{s}}\}_{SU(2)}\ ,\label{sgtssbwyz}\\
\dlt_{\!\!\trm{ns}} \phi & = \{ \phi,G_{\!\!\trm{ns}}\}_{SU(2)},\ \dlt_{\!\!\trm{ns}} \phi^{0}  = \{ \phi^{0},G_{\!\!\trm{ns}}\}_{SU(2)} \ .\label{nsgtssbwyz}
\end{align}
\end{subequations}
The corresponding variations for $ W^{\bar{a}}_{\mu} $, $ Y_{\mu} $ and $ Z_{\mu} $ are given in Eqs.~\eqref{gtcassu2}. Since the gauge algebra Eq.~\eqref{gasu2ssb} is isomorphic to  Eq.~\eqref{gasu2}, it follows that on the constraint surface the algebra of SGTs and NSGTs also becomes Eq.~\eqref{SGT-NSGT alg}. The finite version of SGTs corresponds to the action of $ SU(2) $, whereas the NSGTs are associated with broken generators.

In this subsection we have recast an $ SU(3) $ manifestly invariant theory as  an $ SU(2) $ manifestly invariant theory, \cf~Eqs.\eqref{LS3} and  \eqref{L2SSB} via the admissible point transformation, Eqs.~\eqref{pt1} and \eqref{ptfi}. In the context of theories with SSB, it is said that the $SU(2)$ symmetry is exact, whereas the $SU(3)$ is hidden. We now turn to discuss the SSB of the $SU(3)$ group into the $ SU(2) $ one, from the viewpoint of the latter.

\subsection{SSB from the {$SU(2)$}  perspective}

We reconsider the case of infinite degeneracy of vacuum, $ \mu^{2}<0 $. Configurations with minimal energy Eq.~\eqref{minssb} lie on  $ \phi_{\!\!\trm{min}}^{\dag}\phi_{\!\!\trm{min}} +\phi^{0\, *}_{\!\!\trm{min}}\phi^{0}_{\!\!\trm{min}}= v^{2}$. As we have remarked, there is a natural separation of $ SU(3) $ parameters into those parameters of the isotropy group, $ \al^{\bar{a}} $, and those associated to the broken part of the group, $ \al^{\hat{a}} $ and $ \al^{8} $. In fact, this split is what determines the SGTs and NSGTs previously defined. The functional form of the Lagrangian Eq.~\eqref{L2SSB}, where the $ SU(2) $ sector of $ SU(3) $ is manifest, suggests the study of the following cases:
\begin{itemize}
\item[$  (i)$]\emph{The Goldstone theorem}. We assume the broken part of $ SU(3) $, generated by $ \frac{\lm^{\hat{a}}}{2} $ and $ \frac{\lm^{8}}{2} $,  to be global --that is, we allow $ \al^{\hat{a}} $ and $ \al^{8} $ to be spacetime independent. In other words, assume that the NSGTs are global, but not necessarily SGTs. In such a situation, the following Lagrangian is invariant under this class of transformations:
\begin{equation*}
{\mc L}_g=-\frac{1}{4}W^{\bar{a}}_{\mu \nu}W^{\mu \nu}_{\bar{a}}+(D_\mu \phi)^\dag (D^\mu \phi)+(\partial_{\mu}\phi^{0*})(\partial^\mu \phi^0)+V|_{\Phi \to \phi} \ ,
\end{equation*}
where $ W^{\bar{a}}_{\mu\nu} $ are the components of the $ su(2)$-valued curvature and $ D_{\mu} $ is the covariant derivative  of $ SU(2) $ in the fundamental representation. There arise five massless scalars when the theory is developed around the particular minimum $ \Phi_{\!\!\trm{min}} $, which is decomposed into the doublet $ \phi_{\!\!\trm{min}}=0 $ and the scalar $ \phi^{0}_{\!\!\trm{min}}=v $, by carrying out the shift $ \phi^{0}\mapsto H+i\phi_{Z}\equiv\phi^{0}-v $. These scalars do correspond to $ \phi $, $ \phi^{\dag} $ and the singlet $ \phi_{Z} $, which are identified with the so-called Goldstone bosons. The massive field $ H $ survives. Hence, there is a massless scalar associated with each independent NSGT.

\item[$ (ii) $]\emph{The Higgs mechanism}. Now assume the larger symmetry $ SU(3) $ --that is, that both the SGTs and NSGTs are local.  In this scenario, the theory developed around the particular minimum is characterized by the Lagrangian given in Eq. (\ref{L2SSB}), with $\phi^0$ replaced by $(v+H+i\phi_Z)$.
Five gauge fields, $ Y_{\mu} $, $ Y_{\mu}^{\dag} $, and $ Z_{\mu} $, acquire mass and simultaneously five pseudo-Goldstone bosons appear, namely $ \phi $, $ \phi^{\dag} $ and $ \phi_{Z} $. Notice that all the mass terms are invariant under the $ SU(2) $ subgroup.

All pseudo-Goldstone bosons can be removed from the theory through the so-called unitary gauge; the degrees of freedom that they represent appear as the longitudinal polarization states of the gauge bosons associated with the broken generators.
The implementation of the unitary gauge can be understood in terms of the NSGTs. Indeed, consider the NSGT \eqref{nsgtfi} with particular gauge parameters
\begin{subequations}\label{ug}
\begin{align}
\bt & =  -\dfrac{i\sqrt{2}}{v}\phi\ , \label{ugbt}\\
\al_{Z} & =  -\dfrac{\sqrt{3}}{v} \phi_{Z}\ ,  \label{ugal}
\end{align}
\end{subequations}
which yields  $\phi'=0$ and $\phi'_Z=0$.  Therefore, \emph{the unitary gauge corresponds to a particular NSGT which maps the pseudo Goldstone bosons onto zero}. In addition, from the NSGT given by Eqs.~(\ref{nsgtsu2}), one finds
    \begin{subequations}\label{uggf}
    \begin{align}
    W'^{\bar{a}}_\mu &= W^{\bar{a}}_\mu \ ,  \label{uggf1}\\
    Y'_\mu &= Y_\mu-\frac{i\sqrt{2}}{v}\partial_\mu \phi \ , \label{uggf2}\\
     Z'_\mu &= Z_\mu-\frac{\sqrt{3}}{v}\partial_\mu \phi_Z \  .\label{uggf3}
    \end{align}
    \end{subequations}
    The incorporation of the pseudo Goldstone bosons as the longitudinal component of the massive gauge bosons $Y'_\mu$ and $Z'_\mu$ is evident from these expressions. We will come back to this latter on, when discussing this mechanism in the context of theories with compactified extra dimensions.

\end{itemize}

The unitary gauge can also be implemented via a finite NSGT. Consider the non-linear parametrization of the triplet $\Phi$,
\begin{equation}
  \Phi(x)=\textbf{U}(x)
  \begin{pmatrix}
		0  \\
		0  \\
		{v+H}
	\end{pmatrix}\ ,
    \end{equation}
with
\begin{eqnarray}
\textbf{U}(x)&=&\exp\left(i\frac{\lambda^{\hat{a}}}{2}\al^{\hat{a}}+i\frac{\lambda^{8}}{2}\al^{8}\right) \nonumber \\
&=&\exp\Big\{ -\left(\frac{i}{2v}\right)\Big[i\lambda^4\left(\phi^1-\phi^{1*}\right)-\lambda^5\left(\phi^1+\phi^{1*}\right)\nonumber \\
&&+i\lambda^6\left(\phi^2-\phi^{2*}\right)-\lambda^7\left(\phi^2+\phi^{2*}\right)+\sqrt{\frac{3}{2}}\lambda^8 \phi_Z\Big]\Big\}\, ,
\end{eqnarray}
where the parameter values given in \eqref{ug} were used. The finite version of the NSGT  \eqref{nsgtfi}  are obtained by acting with $\textbf{U}^{-1}(x)$ as follows:
\begin{equation}
\Phi'(x)=\textbf{U}^{-1}(x)\Phi=
\begin{pmatrix}
0 \\
0 \\
{v+H}
\end{pmatrix} \, .
\end{equation}
The components of Eq.~\eqref{uggf}  are recovered by entering the particular element $\textbf{U}^{-1}(x)\in SU(3)$ into the finite gauge transformation of the connection, $ A'_\mu=U(x)A_\mu U^{\dag}(x)-i(\partial_\mu U)U^\dag$,  and keeping the analysis at first order.

\section{Yang-Mills theories with compactified extra dimensions}
\label{YM5}

In this section, we introduce a pure higher-dimensional Yang-Mills theory with an underlying gauge group $ SU(N,\mc{M}^{m}) $, whose parameters are allowed to propagate in the spacetime manifold $ \mc{M}^{m}=\mc{M}^{4}\times \mc{N} ^{n}$. Gauge fields $ \mc{A}^{a}_{M} $, defined on $ \mc{M}^{m} $, act as fundamental fields in the $ m $-dimensional theory, where $ a $ and $ M $ are gauge and spacetime indices, respectively. We begin our discussion by noticing that the transition from the $SU(N,{\mc M}^{m})$ gauge group description to $SU(N,{\mc M}^4)$ will simultaneously convey a certain transformation that maps well defined objects under the Poincar\' e group $ISO(1,m-1)$ onto well defined objects under the standard $ISO(1,3)$. We now proceed to present a brief discussion on this issue.
 
\subsection{The Poincar\' e group perspective}

Let us consider the flat spacetime manifold $ \mc{M}^{m}=\mc{M}^{4}\times \mc{N} ^{n} $, with mostly minus metric $ g_{MN} $ and $ n $ spatial extra dimensions, with coordinates $ (X^{M})=(x^{\mu},x^{\bar{\mu}}) $, where $ \mu=0,1,2,3 $  and $ \bar{\mu}=5,\ldots,m $. We introduce gauge fields $ \mc{A}_{M}(X)=\mc{A}^{a}_{M}(X)T^{a} $, where $ T^{a} $ are generators of the gauge group $ SU(N,\mc{M}^{m}) $. In this $ m $-dimensional spacetime, the Poincar\' e group $ISO(1,m-1)$ is defined through its $\frac{1}{2}m(m+1)$ generators. A number $m$ of these generators ($P_M$) belong to the group of translations, and the $\frac{1}{2}m(m-1)$ remainder ($J_{MN}$) are associated with the Lorentz group $SO(1,m-1)$. These generators satisfy the following Poincar\' e algebra:

\begin{eqnarray}
\label{PL1}
&&[P_M\, ,\, P_N]=0\, , \\
\label{PL2}
&&[J_{MN}\, ,\, P_R]=i\left(g_{MR}P_{N}-g_{NR}P_M\right)\, , \\
\label{PL3}
&&[J_{MN}\, , \, J_{RS}]=i\left(g_{MR}J_{NS}-g_{MS}J_{NR}-g_{NR}J_{MS}+g_{NS}J_{MR}\right)\, .
\end{eqnarray}
It is not difficult to see that in this algebra there are two subalgebras merged. One of these algebras generates the Poincar\' e group $ISO(1,3)$:
\begin{eqnarray}
&&[P_\mu \, ,\, P_\nu ]=0\, , \\
&&[J_{\mu \nu }\, ,\, P_\rho]=i\left(g_{\mu \rho}P_{\nu }-g_{\nu \rho}P_\mu \right)\, , \\
&&[J_{\mu \nu}\, , \, J_{\rho \sigma}]=i\left(g_{\mu \rho}J_{\nu \sigma}-g_{\mu \sigma}J_{\nu \rho}-g_{\nu \rho}J_{\mu \sigma}+g_{\nu \sigma}J_{\mu \rho}\right)\, ,
\end{eqnarray}
whereas the other one generates the inhomogeneous orthogonal group in $n$ dimensions $ISO(n)$:
\begin{eqnarray}
&&[P_{\bar{\mu}}\, , \,P_{\bar{\nu}}]=0\, , \\
&&[J_{\bar{\mu}\bar{\nu}}\, , \,P_{\bar{\rho}}]=i\left(\delta_{\bar{\nu}\bar{\rho}}P_{\bar{\mu}}-\delta_{\bar{\mu}\bar{\rho}}P_{\bar{\nu}}\right)\, ,\\
&&[J_{\bar{\mu}\bar{\nu}}\, , \, J_{\bar{\rho}\bar{\sigma}}]=i\left( \delta_{\bar{\mu}\bar{\sigma}}J_{\bar{\nu}\bar{\rho}}-\delta_{\bar{\mu}\bar{\rho}}J_{\bar{\nu}\bar{\sigma}}-\delta_{\bar{\nu}\bar{\sigma}}J_{\bar{\mu}\bar{\rho}}+
\delta_{\bar{\nu}\bar{\rho}}J_{\bar{\mu}\bar{\sigma}}\right) \, .
\end{eqnarray}

An infinitesimal Poincar\' e transformation in $ {\mc M}^{m}$ is given by
\begin{equation}
\dlt X^M=\omega^{MN}X_N+\epsilon^M \, ,
\end{equation}
where $\omega^{MN}=-\omega^{NM}$ and $\epsilon^M$ are the infinitesimal parameters of the group. This transformation induces the following variation:
\begin{equation}
\delta{\mc A}_M(X)=\left[\omega_{MN}+g_{MN}\left(\omega_{RS}X^S+\epsilon_R\right)\partial^R\right]{\mc A}^N(X) \, .
\end{equation}
This relation can be naturally split into variations for ${\mc A}_\mu(X)$ and ${\mc A}_{\bar{\mu}}(X)$ components as follows:
\begin{subequations}\label{12}
\begin{align}
\delta{\mc A}_\mu(X) =& \ \left[\omega_{\mu \nu}+g_{\mu \nu}\left(\omega_{\ro\sgm}x^\sgm+\epsilon_\ro \right)\partial^\ro\right]{\mc A}^\nu(X) \nonumber\\
&\ +\left[\left(\omega_{\bar{\ro}\bar{\sgm}}x^{\bar{\sgm}}+\epsilon_{\bar{\ro}}\right)\partial^{\bar{\ro}}+\omega_{\ro \bar{\sgm}}\left(x^{\bar{\sgm}}\partial^\ro-x^\ro \partial^{\bar{\sgm}}\right)\right]{\mc A}_\mu(X) \nonumber \\
& \ +\omega_{\mu \bar{\nu}}{\mc A}^{\bar{\nu}}(X) \label{1}\ ,\\
\delta{\mc A}_{\bar{\mu}}(X)=&\ \left[\omega_{\bar{\mu} \bar{\nu}}+g_{\bar{\mu} \bar{\nu}}\left(\omega_{\bar{\ro} \bar{\sgm}}x^{\bar{\sgm}}+\epsilon_{\bar{\ro}} \right)\partial^{\bar{\ro}}\right]{\mc A}^{\bar{\nu}}(X) \nonumber \\
&\ +\left[\left(\omega_{\ro\sgm}x^{\sgm}+\epsilon_{\ro}\right)\partial^{\ro}+\omega_{\ro \bar{\sgm}}\left(x^{\bar{\sgm}}\partial^\ro-x^\ro \partial^{\bar{\sgm}}\right)\right]{\mc A}_{\bar{\mu}}(X) \nonumber \\
&\ +\omega_{\bar{\mu} \nu}{\mc A}^{\nu}(X) \label{2}\, .
\end{align}
\end{subequations}
It can be seen from these expressions that ${\mc A}_{\mu}$ and ${\mc A}_{\bar{\mu}}$ transform under the Lorentz group $SO(1,3)$ as a vector and as a scalar, respectively, whereas they transform as a scalar and as a vector under the orthogonal group $SO(n)$. This means that before compactification, the $ m $-dimensional Yang-Mills action $S[{\mc A}_M]$ (manifestly invariant under $ISO(1,m-1)$) can be written in terms of well defined objects under $ISO(1,3)$ and $ISO(n)$. Thus we can recast this theory in terms of the action $S[{\mc A}_\mu, {\mc A}_{\bar{\mu}}]$. In the latter formulation the $ISO(1,3)$ and $ISO(n)$ symmetries are manifest, but the $ISO(1,m-1)$ is hidden. In complete analogy with the ideas introduced in previous sections for unitary gauge groups, we can define two types of standard transformations, which correspond to the inhomogeneous subgroups $ISO(1,3)$ and $ISO(n)$. The former, which we will call standard Poincar\' e transformations (SPTs), are defined by setting $\omega_{\bar{\mu}\bar{\nu}}=\omega_{\mu\bar{\nu}}=\epsilon_{\bar{\mu}}=0$ in Eqs.~\eqref{12}:
\begin{subequations}
\begin{align}
\delta{\mc A}_\mu(X)&=\left[\omega_{\mu \nu}+g_{\mu \nu}\left(\omega_{\ro \sgm}x^\sgm+\epsilon_\ro \right)\partial^\ro\right]{\mc A}^\nu(X)\ , \\
\delta{\mc A}_{\bar{\mu}}(X)&=\left(\omega_{\ro\sgm}x^{\sgm}+\epsilon_{\ro}\right)\partial^{\ro}{\mc A}_{\bar{\mu}}(X) \ .
\end{align}
\end{subequations}
The latter ones, which we will call standard orthogonal transformations (SOTs), arise when $\omega_{\mu \nu}=\omega_{\mu\bar{\nu}}=\epsilon_{\mu}=0$ in Eqs.~\eqref{12}:
\begin{subequations}
\begin{align}
\delta{\mc A}_\mu(X)&=\left(\omega_{\bar{\ro}\bar{\sgm}}x^{\bar{\sgm}}+\epsilon_{\bar{\ro}}\right)\partial^{\bar{\ro}}{\mc A}_\mu(X)\ , \\
\delta{\mc A}_{\bar{\mu}}(X)&=\left[\omega_{\bar{\mu} \bar{\nu}}+g_{\bar{\mu} \bar{\nu}}\left(\omega_{\bar{\ro} \bar{\sgm}}x^{\bar{\sgm}}+\epsilon_{\bar{\ro}} \right)\partial^{\bar{\ro}}\right]{\mc A}^{\bar{\nu}}(X) \ .
\end{align}
\end{subequations}

The action $S[{\mc A}_\mu, {\mc A}_{\bar{\mu}}]$ is manifestly invariant under these standard spacetime transformations. However, this action is not manifestly invariant under transformations induced by the $J_{\mu \bar{\nu}}$ generators. These are nonstandard Poincar\' e transformations (NSPTs), which are defined from \eqref{12} by setting the parameters $\omega_{\mu \bar{\nu}}\neq 0$ and the remaining ones equal to zero:
\begin{subequations}
\begin{align}
\delta{\mc A}_\mu(X) =&\omega_{\ro \bar{\sgm}}\left(x^{\bar{\sgm}}\partial^\ro-x^\ro \partial^{\bar{\sgm}}\right){\mc A}_\mu(X)+\omega_{\mu \bar{\nu}}{\mc A}^{\bar{\nu}}(X) \ , \\
\delta{\mc A}_{\bar{\mu}}(X) =&\omega_{\ro \bar{\sgm}}\left(x^{\bar{\sgm}}\partial^\ro-x^\ro \partial^{\bar{\sgm}}\right){\mc A}_{\bar{\mu}}(X)+\omega_{\bar{\mu} \nu}{\mc A}^{\nu}(X) \ .
\end{align}
\end{subequations}

In the  five-dimensional pure Yang-Mills theory with one compact spatial extra dimension, there arise massless bosons that are interpreted as pseudo-Goldstone bosons. These fields can be removed via a particular NSGT which is understood as a unitary gauge~\citte{NT}. Although these pseudo-Goldstone bosons are present, in  the switch from the gauge group $SU(N,{\mc M}^{5})$ to  $SU(N,{\mc M}^4)$ there is no SSB involved, because the number of generators in both groups is the same. So, in this class of theories the pseudo-Goldstone bosons needed to implement the Higgs mechanism have nothing to do with the unitary gauge group $SU(N,{\mc M}^{5})$, but with the Poincar\' e group. The boson fields arise by compactification of the spatial extra coordinates which leads to an explicit breaking of the $ISO(1,4)$ group into $ISO(1,3)$. This observation implies that the corresponding effective theory, which depends on the KK fields, is subject to satisfying only the SPTs. We expect a similar behavior when considering compactification of  higher-dimensional pure $ SU(N,\mc{M}^{m}) $ Yang-Mills theories into $ SU(N,\mc{M}^{4}) $ effective theory.

\subsection{Pure {$ SU(N,\mc{M}^{m}) $} Yang-Mills Theory}
The Lagrangian that describes pure $ SU(N,\mc{M}^{m}) $ Yang-Mills theory is given by (\cf \eqref{L3})
 \begin{equation}\label{LYMm}
 {\mc L}_{ SU\!(N,\,\mc{M}) }(x,y)=-\frac{1}{4}{\mc F}^a_{MN}(x,y){\mc F}^{MN}_a(x,y)\ ,
 \end{equation}
where in this subsection $ (x,y) $ denotes the coordinates of $ \mc{M}^{4}\times \mc{N}^{n} $. The components $ \mc{F}^a_{MN} $ are regarded as functions of  gauge fields $ {\mc A}^{a}_{M}(x,y) $ as in Eq.~\eqref{fmn3} except that in this case the coupling constant  is denoted by $ g_{m} $, whose dimension is of  [mass]${}^{(4-m)/2}  $. Gauge invariances of this theory are (\cf  \eqref{gt})
\begin{equation}\label{gtm}
 \delta {\mc A}^a_{M}={\mc D}^{ab}_M \alpha^b(x,y)\ ,
\end{equation}
where ${\mc D}^{ab}_M=\delta^{ab}\partial_M-g_mf^{abc}{\mc A}^c_M$ and the gauge parameters are allowed to propagate in the bulk. From Eq.~\eqref{gtm}, the  components of the curvature are transformed in the adjoint representation $\delta {\mc F}^a_{MN}=g_mf^{abc}{\mc F}^b_{MN}\alpha^c(x,y)$ .

The Hamiltonian description of the theory goes along the same line as Sect. \ref{secYM3}. The conjugate momentum to $ \mc{A}_{M}^{a} $ is denoted by $ \pi^{M}_{a} $. The canonical analysis yields the following first-class constraints:
\begin{subequations}\label{fccmm}
\begin{align}
\phi^{(1)}_{a} & =\pi^{0}_{a}(x,y)\approx 0  \label{pcm}\\
\phi^{(2)}_{a} & =\mc{D}_{I}^{ab}\pi^{I}_{b}(x,y)\approx 0 \label{scm}
\end{align}
\end{subequations}
where $ I $ labels all spatial components of $ \mc{M}^{m} $. Therefore, the number of physical degrees of freedom is $ (N^{2}-1)m-2(N^{2}-1)=(N^{2}-1)(m-2)$ per spatial point of  $ \mc{M}^{m}$.

The corresponding gauge algebra has the structure of Eq.~\eqref{la3} with the corresponding coupling constant $ g_{m} $:
\begin{equation}\label{gaym5}
\{ \phi^{(2)}_a[u],\phi^{(2)}_b[v]\}_{SU(N,\mc{M})}=g_{m}f_{abc}\,\phi^{(2)}_c[uv]\ ,
\end{equation}
where the Poisson bracket $ \{\cdot,\cdot\}_{SU\!(N,\mc{M})} $ is calculated in terms of canonical conjugate pairs $ (\mc{A}_{M}^{a},\pi^{M}_{a}) $. In the same fashion, gauge transformations Eq.~\eqref{gtm} can be obtained  via the corresponding gauge generator \cf~Eq.~\eqref{gg3} as follows:
\begin{equation}\label{gtcam}
\dlt A_{M}^{a}= \{\mc{A}^{a}_{M},G\}_{SU\!(N,\mc{M})}\ .
\end{equation}

We now perform the transition from the $ SU(N,\mc{M}^{m}) $ variables to the natural variables that arise in the effective theory after compactification.

\subsection{Compactified theory and the  {$ SU(N,\mc{M}^{4}) $} description}
For the sake of simplicity,  from now on we focus on the case $ n=1 $; that is, the five-dimensional $ SU(N,\mc{M}^{5}) $ Yang-Mills theory. The notion of hidden symmetry induced by a canonical transformation will be given in terms of Fourier transformations and the identification of $ G $ as $ SU(N,\mc{M}^{5}) $ and $ H $ as $ SU(N,\mc{M}^{4}) $. In five dimensions, the theory consists of $ 3(N^{2}-1) $ true degrees of freedom per spatial point of $ \mc{M}^{5}$.

The components $ \mc{A}_{M}^{a}(x,y) $ of the connection  find a natural split into $ \mc{A}^{a}_{\mu}(x,y) $ and $ \mc{A}^{a}_{5}(x,y) $, and  following Ref.~\cite{NT}, we assume the compact extra dimension homotopically equivalent to the circle $ S^{1}$ of radius $ R $. Fields $\mc{A}^{a}_{\mu}(x,y) $ and $ \mc{A}^{a}_{5}(x,y) $ are assumed to be periodic with respect to the fifth coordinate, so they can be expressed as Fourier series. 
In order to recover a pure four-dimensional Yang-Mills sector within the effective theory, we introduce a further symmetry in the compact extra dimension by replacing it with $ S^{1}/Z_{2} $, hence $y$ is identified with $-y$. We assume that $\mc{A}^{a}_{\mu}(x,y) $ and $ \mc{A}^{a}_{5}(x,y) $ are, respectively, even and odd under the reflection $y\to -y$; these imply that curvature components $ \mc{F}_{\mu\nu}^{a}(x,y) $ and $ \mc{F}_{\mu 5}^{a}(x,y) $ display even and odd parity in the extra dimension, respectively. Under these assumptions, the following Fourier expansions are allowed:
\begin{subequations}\label{pt5}
\begin{align}
\imc{A}{a}{\mu}(x,y)& = \frac{1}{\sqrt{R}}{A}^{(0)a}_{\mu}(x)+\sqrt{\frac{2}{R}}\sum_{m=1}^{\infty}{A}^{(m)a}_{\mu}(x)\cos \left(2\pi\frac{my}{R}\right)\ , \label{pt5amu}\\
\imc{A}{a}{5}(x,y)&= \sqrt{\frac{2}{R}}\sum_{m=1}^{\infty}{A}^{(m)a}_{5}(x)\sin \left(2\pi\frac{my}{R}\right)\ ,\label{pt5a5}\\
\imc{F}{a}{\mu\nu}(x,y) & = \frac{1}{\sqrt{R}}\imc{F}{(0)a}{\mu\nu}(x)+\sqrt{\frac{2}{R}}\sum_{m=1}^{\infty}\imc{F}{(m)a}{\mu\nu}(x)\cos \left(2\pi\frac{my}{R}\right) \ ,\label{pt5fmn}\\
\imc{F}{a}{\mu 5}(x,y)&= \sqrt{\frac{2}{R}}\sum_{m=1}^{\infty}\imc{F}{(m)a}{\mu 5}(x)\sin \left(2\pi\frac{my}{R}\right)\ .\label{pt5fm5}
\end{align}
\end{subequations}
In particular, it will be important to make the analogy between Eqs. \eqref{pt5amu} and \eqref{pt5a5} and the point transformations in Eq.~\eqref{pt1}.

Following the compactification scheme introduced in Ref.~\cite{NT}, one obtains the Fourier components of the curvature in terms of the gauge fields Fourier modes:
\begin{subequations}\label{fmnfourier}
\begin{align}
{\mc F}^{(0)a}_{\mu \nu} & =F^{(0)a}_{\mu \nu}+gf^{abc}A^{(m)b}_\mu A^{(m)c}_\nu \  ,\label{fmn-0}\\
{\mc F}^{(m)a}_{\mu \nu} & ={\mc D}^{(0)ab}_\mu A^{(m)b}_\nu-{\mc D}^{(0)ab}_\nu A^{(m)b}_\mu +gf^{abc}\Delta_{mrn}A^{(r)b}_\mu A^{(n)c}_\nu \ ,\label{fmn-m}\\
{\mc F}^{(m)a}_{\mu 5} & ={\mc D}^{(0)ab}_\mu A^{(m)b}_5+\frac{2 \pi m}{R}A^{(m)a}_\mu +gf^{abc}\Delta'_{mnr}A^{(r)b}_\mu A^{(n)c}_5 \ ,\label{fm5-m}
\end{align}
\end{subequations}
where ${\mc D}^{(0)ab}_\mu=\delta^{ab}\partial_\mu-gf^{abc}A^{(0)c}_\mu$, the coupling constant $g=g_5/\sqrt{R}$, and
\begin{equation}\label{Fmn-0}
F^{(0)a}_{\mu \nu}=\partial_\mu A^{(0)a}_\nu-\partial_\nu A^{(0)a}_\mu+gf^{abc}A^{(0)b}_\mu A^{(0)c}_\nu \ .
\end{equation}
In addition
\begin{subequations}\label{dlts}
\begin{align}
\Delta_{mrn} & =\frac{1}{\sqrt{2}}\left(\delta_{r,m+n} +\delta_{m,r+n}+\delta_{n,r+m}\right)\  , \label{dlt}\\
\Delta'_{mrn} & =\frac{1}{\sqrt{2}}\left(\delta_{m,r+n} +\delta_{r,m+n}-\delta_{n,r+m}\right) \ .\label{dltprime}
\end{align}
\end{subequations}
Notice that there is a clear resemblance between Eqs.~\eqref{fmn2} and \eqref{fmnfourier}. In the same fashion that the $ su(3) $-valued curvature in our toy model  was decomposed into well defined objects ($ F^{\bar{a}}_{\mu\nu} $, $ Y_{\mu\nu} $, and $ F^{8}_{\mu\nu} $) under the $ SU(2) $ subgroup, we will show that  the components of Eq.~\eqref{fmnfourier} represent the decomposition of the pure $ SU(N,\mc{M}^{5}) $ Yang-Mills curvature into well defined objects ($ \mc{F}^{(0)a}_{\mu\nu} $, $ \mc{F}^{(m)a}_{\mu\nu} $, and $ \mc{F}^{(0)a}_{\mu 5} $) under the subgroup $ SU(N,\mc{M}^{4}) $. In our toy model, such decomposition was performed  by means of the point transformation in Eq.~\eqref{pt1}; in the present case we will take advantage of Eqs.~\eqref{sgtfmn4}. Moreover, in the present theory, the curvature decomposition is also a map from well defined objects under $ISO(1,4)$ onto well defined objects under $ISO(1,3)$.

Integrating out the extra dimension after Fourier expanding Eq.~\eqref{LYMm} yields the following effective Lagrangian, \cf  \eqref{L2}:
\begin{equation}\label{LYM4}
{\mc L}_{ SU(N,\,\mc{M}^{4}) }=-\frac{1}{4}\left({\mc F}^{(0)a}_{\mu \nu}{\mc F}^{(0)a\mu \nu}+{\mc F}^{(m)a}_{\mu \nu}{\mc F}^{(m)a\mu \nu}+2\, {\mc F}^{(m)a}_{\mu 5}{\mc F}^{(m)a\mu 5}\right)\ .
\end{equation}

The analysis of the toy model in Sec.~\ref{TM} suggests that Fourier expansions of gauge fields, Eqs.~\eqref{pt5amu}  and \eqref{pt5a5}, can be treated as a point transformation which connects the natural coordinates in the pure  five-dimensional Yang-Mills theory $ (\imc{A}{a}{M}) $ and the built-in coordinates $ ({A}^{(0)a}_{\mu}$, $  {A}^{(m)a}_{\mu}$, and ${A}^{(m)a}_{5}  $) of the effective Lagrangian Eq.~\eqref{LYM4}. In this framework, gauge transformations Eq.~\eqref{gtm} are mapped by Eqs.~\eqref{pt5amu}  and \eqref{pt5a5} onto
\begin{subequations}\label{gtgf4}
\begin{align}
\delta A^{(0)a}_\mu & = {\mc D}^{(0)ab}_\mu \alpha^{(0)b}+gf^{abc}A^{(m)b}_\mu \alpha^{(m)c} \  ,\label{gtA0a}\\
\delta A^{(m)a}_\mu &= gf^{abc}A^{(m)b}_\mu \alpha^{(0)c}+{\mc D}^{(mn)ab}_\mu \alpha^{(n)b} \  ,\label{gtAma} \\
\delta A^{(m)a}_5 &= gf^{abc}A^{(m)b}_5\alpha^{(0)c}+{\mc D}^{(mn)ab}_{5}\alpha^{(n)b}\ , \label{gtA5a}
\end{align}
\end{subequations}
after the extra dimension is integrated out. The parameters $ \al^{(0)a}(x) $ and $ \al^{(m)a}(x) $ are the Fourier components in the expansion of $ \al^{a}(x,y)=\al^{a}(x,-y) $.  In Eq.~\eqref{gtgf4} the following quantities have been defined:
\begin{subequations}\label{Ds}
\begin{align}
	{\mc D}^{(mn)ab}_\mu &= \delta^{mn}{\mc D}^{(0)ab}_\mu-gf^{abc}\Delta_{mrn}A^{(r)c}_\mu\ ,  \label{Dmnmu}\\
	{\mc D}^{(mn)ab}_5 & = -\frac{2\pi m}{R}\delta^{mn}\delta^{ab}-gf^{abc}\Delta'_{mrn}A^{(r)c}_5\ . \label{Dmn5}
\end{align}
\end{subequations}

In analogy with Eqs.~\eqref{sgt2} and  \eqref{nsgtsu2}, the SGTs and NSGTs are defined in this case. The SGTs correspond to Eq.~\eqref{gtgf4} after setting $ \al^{(n)a}=0 $:
\begin{subequations}\label{sgtgf4}
\begin{align}
\delta_{\!\!\trm{s}} A^{(0)a}_\mu & = {\mc D}^{(0)ab}_\mu \alpha^{(0)b},\label{sgtA0a}\\
\delta_{\!\!\trm{s}} A^{(m)a}_\mu &= gf^{abc}A^{(m)b}_\mu \alpha^{(0)c} \  ,\label{sgtAma} \\
\delta_{\!\!\trm{s}} A^{(m)a}_5 &= gf^{abc}A^{(m)b}_5\alpha^{(0)c}\ . \label{sgtA5a}
\end{align}
\end{subequations}
In analogy with the gauge fields $ W^{\bar{a}}_{\mu} $ under $ SU(2) $ Eq.~\eqref{sgtW},  the Fourier component $ A^{(0)a}_\mu $ becomes a gauge field with respect to $ SU(N,\mc{M}^{4}) $. Similarly, the matter field $ Y_{\mu} $ is comparable with the excited KK modes $ A^{(n)a}_{\mu} $, which transform in the adjoint representation of $ SU(N,\mc{M}^{4}) $. In addition, $ A^{(n)a}_{5} $ transform as matter fields in the adjoint representation of $ SU(N,\mc{M}^{4}) $. The NSGTs are obtained from Eq.~\eqref{gtgf4} by setting $ \al^{(0)a}\equiv 0 $, that is (\cf~\eqref{nsgtsu2})
\begin{subequations}\label{nsgtgf4}
\begin{align}
\delta_{\!\!\trm{ns}} A^{(0)a}_\mu & = gf^{abc}A^{(m)b}_\mu \alpha^{(m)c} \  ,\label{nsgtA0a}\\
\delta_{\!\!\trm{ns}} A^{(m)a}_\mu &= {\mc D}^{(mn)ab}_\mu \alpha^{(n)b} \  ,\label{nsgtAma} \\
\delta_{\!\!\trm{ns}} A^{(m)a}_5 &= {\mc D}^{(mn)ab}_{5}\alpha^{(n)b}\ . \label{nsgtA5a}
\end{align}
\end{subequations}

Gauge invariance of Eq.~\eqref{LYM4} under Eq.~\eqref{gtgf4} is guaranteed, since the latter imply the following variations at the level of the Fourier components of the curvature:
\begin{subequations}\label{gtfmn4}
\begin{align}
\delta {\mc F}^{(0)a}_{\mu \nu} & =gf^{abc}\left({\mc F}^{(0)b}_{\mu \nu}\alpha^{(0)c}+{\mc F}^{(m)b}_{\mu \nu}\alpha^{(m)c}\right)\ ,
 \label{gtfmn-0}\\
\delta {\mc F}^{(m)a}_{\mu \nu} & =gf^{abc}\left({\mc F}^{(m)b}_{\mu \nu}\alpha^{(0)c}+\left(\delta_{mn}{\mc F}^{(0)b}_{\mu \nu}+\Delta_{mrn}{\mc F}^{(r)b}_{\mu \nu} \right)\alpha^{(n)c}\right)\ , \label{gtfmn-m}\\
\delta {\mc F}^{(m)a}_{\mu 5} & =gf^{abc}\left({\mc F}^{(m)b}_{\mu 5}\alpha^{(0)c}+\Delta'_{mrn}{\mc F}^{(r)b}_{\mu 5}\alpha^{(n)c}\right) \ .
 \label{gtfmn-5}
\end{align}
\end{subequations}
It is not difficult to see that  the effective Lagrangian $ \mc{L}_{SU(N,\,\mc{M}^{4})} $ is invariant under these transformations. Therefore, the components of Eq.~\eqref{gtgf4} are genuine gauge transformations of the effective theory.

It is worth noticing that the scalar fields $A^{(m)a}_5$ can be eliminated altogether via a particular NSGT. Consider a NSGT with infinitesimal gauge parameters given by $\alpha^{(m)a}(x)=(R/2\pi m)A^{(m)a}_5(x)$,\ Ref.~\cite{NT}. Then, from Eq.~\eqref{nsgtA5a}, we can see that $A^{(m)a}_5 \to A'^{(m)a}_5=0$ at first order. This result shows that the $A^{(m)a}_5(x)$ scalar fields are in fact pseudo Goldstone bosons.

It is important to stress that the invariance of the effective theory  Eq.~\eqref{LYM4} under the transformations Eq.~\eqref{gtgf4} is by no means immediate. A direct calculation of the curvature variations Eq.~\eqref{gtfmn4} from  Eq.~\eqref{gtgf4} gives raise to the following extra terms quadratic in $ g $:
\begin{subequations}
\begin{align}
\Dlt_{\mu\nu}^{(m)a} = & -g^{2}\left[f_{abc}f_{bde}(\dlt_{pq}\dlt_{mn}+\Dlt_{rpq}\Dlt_{rmn})+f_{adb}f_{bce}(\dlt_{nq}\dlt_{mp}+\Dlt_{rnq}\Dlt_{rmp})\right. \label{Dlt}\\
 &  \left. + f_{abe}f_{bcd}(\dlt_{np}\dlt_{mq}+\Dlt_{rnp}\Dlt_{rmq})\right]A^{(p)d}_{\mu}A^{(q)e}_{\nu}\al^{(n)c} \ ,\nonumber\\
\wt{\Dlt}_{\mu 5}^{(m)a} = & -g^{2}\left[f_{abc}f_{bde}\Dlt'_{rqp}\Dlt'_{rmn}+f_{adb}f_{bce}\Dlt'_{rqn}\Dlt'_{rmp}\right.\nonumber\\
 &  \left. + f_{abe}f_{bcd}(\dlt_{np}\dlt_{mq}+\Dlt_{npr}\Dlt'_{mqr})\right]A^{(p)d}_{\mu}A^{(q)e}_{5}\al^{(n)c} \ ,\label{wtDlt}
\end{align}
\end{subequations}
in Eqs. \eqref{gtfmn-m} and \eqref{gtfmn-5}, respectively. These terms, that would destroy  the invariance of the effective Lagrangian $ \mc{L}_{SU(N,\mc{M}^{4})} $ under Eq.~\eqref{gtgf4},  are necessarily zero by consistency with the Fourier transformation in Eq.~\eqref{pt5}. The variation of curvatures $ \dlt\mc{F}_{MN}^{a}=g_{5}f_{abc}\mc{F}_{MN}^{b}\al^{c} $ is duly mapped onto Eqs.~\eqref{gtfmn4} under the point transformation Eq.~\eqref{pt5}. We will discuss further this point within the Hamiltonian formalism of the theory.

The SGTs Eq.~\eqref{sgtgf4} induce the corresponding transformations at the curvature level. From Eq.~\eqref{gtfmn4}, all Fourier components of $ \mc{F}_{MN}^{a} $ do  covariantly transform  under the symmetry group of SGTs, $ SU(N,\mc{M}^{4}) $:
\begin{subequations}\label{sgtfmn4}
\begin{align}
\delta_{\!\!\trm{s}} {\mc F}^{(0)a}_{\mu \nu} & =gf^{abc}{\mc F}^{(0)b}_{\mu \nu}\alpha^{(0)c}\ ,
 \label{sgtfmn-0}\\
\delta_{\!\!\trm{s}} {\mc F}^{(m)a}_{\mu \nu} & =gf^{abc}{\mc F}^{(m)b}_{\mu \nu}\alpha^{(0)c}\ , \label{sgtfmn-m}\\
\delta_{\!\!\trm{s}} {\mc F}^{(m)a}_{\mu 5} & =gf^{abc}{\mc F}^{(m)b}_{\mu 5}\alpha^{(0)c} \ .
 \label{sgtfmn-5}
\end{align}
\end{subequations}

The phase space description of this theory allows us to define the gauge generators associated to the so-called SGTs and NSGTs defined above.  The canonical analysis of the effective Lagrangian Eq.~\eqref{LYM4} goes along the same lines of reasoning as Sect.~B2 of Ref.~\cite{NT}. The conjugate momenta are given by
\begin{subequations}\label{pisym4}
\begin{align}
\pi^{(0)\mu} _{a}& = \mc{F}^{(0)\mu0}_{a}\ , \label{pi0mu}\\
\pi^{(n)\mu} _{a}& = \mc{F}^{(n)\mu0}_{a}\ , \label{pinmu}\\
\pi^{(0) 5} _{a}& = \mc{F}^{(n) 50}_{a}\ .  \label{pin5}
\end{align}
\end{subequations}
It is worth noticing, from Eqs.~\eqref{sgtgf4} and \eqref{sgtfmn4}, that canonical pairs are well defined objects with respect to $ SU(N,\mc{M}^{4}) $. In addition, the Fourier expansions in Eqs.~\eqref{pt5fmn} and \eqref{pt5fm5} together with $ \pi^{M}_{a} =\mc{F}^{M0}_{a}$ allow us to write
\begin{subequations}\label{ctym4}
\begin{align}
\pi_{a}^{\mu}(x,y)& = \frac{1}{\sqrt{R}}\pi^{(0)\mu}_{a}(x)+\sqrt{\frac{2}{R}}\sum_{m=1}^{\infty}\pi^{(m)\mu}_{a}(x)\cos \left(2\pi\frac{my}{R}\right)\ , \label{ct5amu}\\
\pi_{a}^{5}(x,y)&= \sqrt{\frac{2}{R}}\sum_{m=1}^{\infty}\pi^{(m)5}_{a}(x)\sin \left(2\pi\frac{my}{R}\right)\ .\label{ct5a5}
\end{align}
\end{subequations}
These expressions relate the conjugate momenta inherent in the pure $ SU(N,\mc{M}^{5}) $ Yang-Mills theory and those present in the effective $ SU(N,\mc{M}^{4}) $ theory. Moreover, they are analogous to Eq.~\eqref{ct}.

The temporal component of Eqs.~\eqref{pi0mu} and \eqref{pinmu} define the following primary constraints:
\begin{subequations}\label{pcym4}
\begin{align}
\phi^{(1)(0)}_{a} & = \pi^{(0)0}_{a}\approx 0 \ ,\label{pcym4-1}\\
\phi^{(1)(n)}_{a} & = \pi^{(n)0}_{a} \approx 0\ .\label{pcym4-2}
\end{align}
\end{subequations}
The primary Hamiltonian takes the form (\cf~\eqref{phsu2})
\begin{equation}\label{phym4}
\mc{H}_{SU(N,\,\mc{M}^{4})}^{(1)}=\mc{H}_{SU(N,\,\mc{M}^{4})} + \mu^{(0)a}\phi^{(1)(0)}_{a}+\mu^{(n)a}\phi^{(1)(n)}_{a}\ ,
\end{equation}
where besides the linear combination of primary constraints, with Lagrange multipliers $\mu^{(0)a}  $ and $ \mu^{(n)a} $ as coefficients, the canonical Hamiltonian is (\cf~\eqref{eq:Hcsu2})
\begin{align}\label{chym4}
\mc{H}_{SU(N,\,\mc{M}^{4})}  = &\  \half\left(\pi^{(0)i}_a\pi^{(0)i}_a+\pi^{(n)i}_a\pi^{(n)i}_a+\pi^{(n)5}_a\pi^{(n)5}_a\right)+\quarter\left(\mc{F}^{(0)ij}_{a}\mc{F}^{(0)a}_{ij}+2
\mc{F}^{(n)i5}_{a}\mc{F}^{(n)a}_{i5}\right) \nonumber \\
&\ -A^{(0)a}_0 \phi^{(2)(0)}_{a}-A^{(n)a}_0 \phi^{(2)(n)}_{a}\ ,
\end{align}
where $ \phi^{(2)(0)}_{a}$ and $\phi^{(2)(n)}_{a} $ are functions of phase space that will be specified after presenting a couple of  key results useful for the rest of the discussion.
\begin{proposition}
\label{prop:ct}
The Fourier expansion of gauge fields and conjugate momenta, Eqs.~\eqref{pt5amu}, \eqref{pt5a5} and \eqref{ctym4}, define a canonical transformation.
\end{proposition}

The proof of this proposition is collected in the Appendix. This proposition ensures that $ \{\cdot,\cdot\}_{SU(N,\mc{M})}=  \{\cdot,\cdot\}_{SU(N,\mc{M}^{4})}$, where $ \{\cdot,\cdot\}_{SU(N,\mc{M}^{4})} $ indicates the Poisson bracket with respect to $ (A^{(0)a}_{\mu},\pi^{(0)\mu}_{a}) $, $ (A^{(n)a}_{\mu},\pi^{(n)\mu}_{a}) $,  and $  (A^{(n)a}_{5},\pi^{(n)5}_{a}) $. Because there exists a spacetime independent canonical transformation between the pure $ SU(N,\mc{M}^{5}) $ Yang-Mills theory and the effective theory based on $ SU(N,\mc{M}^{4}) $, it immediately follows that both canonical Hamiltonians $ \mc{H}_{SU(N,\mc{M})} $ and $ \mc{H}_{SU(N,\mc{M}^{4})} $ are mapped into each other via such canonical transformation, as can be proved by direct calculation. However, in a singular theory, the time evolution is governed by the primary Hamiltonian and not by the canonical one. An important observation is the following: If in a general singular theory of fields there is a spacetime independent canonical transformation which  connects two primary Hamiltonians corresponding to two different formulations of the same theory --that is, if such transformation maps one set of primary constraints into the other one, then both formulations must have the same number of generations of constraints (tertiary, quartic, etc.). This is an immediate consequence of the relation between the Poisson brackets in the two different formulations. Another consequence is that the set of secondary (tertiary, quartic, etc.) constraints in one of the formulations is necessarily mapped onto the corresponding set of constraints in the other formulation via the canonical transformation. The following result allows us to use these observations within the current analysis.
\begin{proposition}\label{prop:pcs}
The set of primary constraints Eq.~\eqref{pcm} of the five dimensional pure $ SU(N,\mc{M}^{5}) $ Yang-Mills theory is faithfully mapped onto the set of primary constraints Eq.~\eqref{pcym4} of the $ SU(N,\mc{M}^{4}) $ Yang-Mills theory.
\end{proposition}

The proof of this proposition is straightforward from Eq.~\eqref{ct5amu} and the linear independence of trigonometric functions. Moreover, it can be extended to the case of $ m $ dimensional pure $ SU(N,\mc{M}^{m}) $ Yang-Mills theory and its compactification down to four dimensions.

Propositions \ref{prop:ct} and \ref{prop:pcs}  ensure that secondary constraints
\begin{subequations}\label{scym4}
\begin{align}
\phi^{(2)(0)}_a &={\mc D}^{(0)ab}_i\pi^{(0)i}_b-gf^{abc}\left( A^{(n)c}_i \pi^{(n)i}_b+A^{(m)c}_5\pi^{(m)5}_b \right)\approx0\ ,\label{scym4-0}\\
\phi^{(2)(n)}_a&= {\mc D}^{(nm)ab}_i\pi^{(m)i}_b-{\mc D}^{(nm)ab}_5\pi^{(m)5}_b-gf^{abc}A^{(n)c}_i\pi^{(0)i}_b \approx 0\ , \label{scym4-n}
\end{align}
\end{subequations}
that emerge in the canonical Hamiltonian Eq.~\eqref{chym4} can be also calculated from Eq.~\eqref{scm} via the canonical transformation mentioned in Prop.~\ref{prop:ct}. Less trivial outcomes of the considerations above are the following: First, the effective theory must not present either tertiary or higher constraint generations. Second, the gauge algebra of the effective theory can be obtained via the canonical transformation from the gauge algebra Eq.~\eqref{gaym5}  of the pure five dimensional Yang-Mills theory. In fact,
\begin{subequations}\label{gaym4}
\begin{align}
\{\phi^{(2)(0)}_a[u],\phi^{(2)(0)}_b[v]\} & =gf_{abc}\phi^{(2)(0)}_c[uv]\ , \label{gaym4-1}\\
\{\phi^{(2)(0)}_a[u],\phi^{(2)(n)}_b[v]\} & =gf_{abc}\phi^{(2)(n)}_c[uv]\  , \label{gaym4-2}\\
\{\phi^{(2)(m)}_a[u],\phi^{(2)(n)}_b[v]\} & =gf_{abc}\left(\delta_{mn}\phi^{(2)(0)}_c[uv]+\Delta_{mnr}\phi^{(2)(r)}_c [uv]\right)\ ,\label{gaym4-3}
\end{align}
\end{subequations}
which coincides with Eqs.~(68)-(70) of Ref.~\cite{NT}.

The gauge generator that reproduces the gauge transformations in Eq.~\eqref{gtgf4} is the sum of the SGTs ($  G_{\!\!\trm{s}} $) plus the NSGTs ($  G_{\!\!\trm{ns}} $) generators, where
\begin{subequations}\label{ggym4}
\begin{align}
 G_{\!\!\trm{s}}& =\left(\mc{D}^{(0)ab}_{0}\al^{(0)b}\right)\phi^{(1)(0)}_{a}+gf_{abc}A^{(n)b}_{0}\al^{(0)c}\phi^{(1)(n)}_{a}-\al^{(0)a}\phi^{(2)(0)}_{a} \label{ggsgt4}\ ,\\
 G_{\!\!\trm{ns}}& =gf_{abc}A^{(m)b}_{0}\al^{(m)c}\phi^{(1)(0)}_{a}+\left(\mc{D}^{(mn)ab}\al^{(n)b}\right)\phi^{(1)(m)}_{a}-\al^{(m)a}\phi^{(2)(m)}_{a}\ .\label{ggnsgt4}
\end{align}
\end{subequations}
From the transformation laws generated by $  G_{\!\!\trm{s}} $ and $ G_{\!\!\trm{ns}} $, together with the constraint algebra Eq.~\eqref{gaym4}, one can infer the Lie algebra Eq.~\eqref{SGT-NSGT alg} on the constraint surface for the SGTs and NSGTs in this case. Due to the constraint algebra Eq.~\eqref{gaym4-1},  the SGTs exponentiate into $ SU(N) $, and since in $ G_{\!\!\trm{s}} $ the gauge parameters $ \al^{(0)a} $ are defined on $ \mc{M}^{4} $, we have that exponentiation of SGTs provides $ SU(N,\mc{M}^{4}) $; the algebra of NSGTs does not close, hence this transformations do not exponentiate into a group.  The sum $  G_{\!\!\trm{s}} +  G_{\!\!\trm{ns}} $ is the image under the canonical transformation mentioned in Prop.~\ref{prop:ct} of the gauge generator that reproduces gauge transformations Eq.~\eqref{gtm} in the five dimensional case.

If a complete set of gauge transformations at the Hamiltonian level can be found, then a complete set of gauge transformations at the Lagrangian level can be recovered \cite{hen90}. This implies that there are no more gauge invariances of the Lagrangian Eq.~\eqref{LYM4} than those altogether generated by Eq.~\eqref{ggym4}, which in turn correspond to  Eq.~\eqref{gtgf4}. Therefore, the effective Lagrangian Eq.~\eqref{LYM4} must be invariant under these transformations, so that any extra term in the calculation of $ \dlt\mc{L}_{SU(N,\mc{M}^{4})} $ must be either identically zero or a surface term. In this regard we argue that the extra terms Eqs.~\eqref{Dlt} and  \eqref{wtDlt} must vanish since they do not include any derivative, hence they cannot be rewritten as a surface term.

We end this section with a heuristic counting of true degrees of freedom in the effective theory. Let us take for the moment ``truncated Fourier expansions" up to some order $ K $, so that, letting $ K\to\infty $ will precisely yield $ (\mc{A}^{a}_{\mu}(x,y),\pi^{\mu}_{a}(x,y)) $ and $ (\mc{A}^{a}_{5}(x,y),\pi^{5}_{a}(x,y)) $ in terms of $ (A^{(0)a}_{\mu}(x),\pi^{(0)\mu}_{a}(x)) $, $ (A^{(n)a}_{\mu}(x),\pi^{(n)\mu}_{a}(x)) $, $ (A^{(n)a}_{5}(x),\pi^{(n)5}_{a}(x)) $ and trigonometric functions. In other words, $ K $ quantifies the contribution from the extra dimension in the ``truncated Fourier expansions".  The number of  canonical pairs and first-class constraints in the truncated version are $ 2\times[4(N^{2}-1)+ 4K(N^{2}-1)+K(N^{2}-1)]$  and $2(N^{2}-1)+2K(N^{2}-1)$, respectively. Thus, the number of true degrees of freedom when $ K $ is large but finite is $N_{0}(K)=2(N^{2}-1)+3K(N^{2}-1) $ per spatial point of $ \mc{M}^{4} $. Allowing $ K\to\infty $ causes this number of true degrees of freedom to diverge, precisely because one is also counting the continuum contribution of the extra dimension. In order to obtain the number of true degrees of freedom per spatial point of $ \mc{M}^{5} $, one needs to take the ratio $ N_{0}/K $ before considering $ K\to\infty $. After this process is done, we have that the number of true degrees of freedom per spatial point of $ \mc{M}^{5} $ is $ 3(N^{2}-1) $, which coincides with the corresponding number in the pure $ SU(N,\mc{M}^{5}) $ Yang-Mills theory.

\section{Final remarks}
\label{FR}
In order to clarify the gauge structure of pure five-dimensional Yang-Mills theories formulated on a spacetime manifold with a compact spatial extra dimension, a notion of hidden symmetry based on the fundamental concept of canonical transformation was introduced.  Although the idea of hidden symmetry is well known in the context of theories with SSB, we have extended this notion to include more general scenarios. The canonical transformation under consideration maps well defined objects under a gauge group $ G $ to well defined objects under a non-trivial subgroup $ H \subset G$. This transformation was constructed within two different categories depending whether the subgroup $ H $ is generated $ (a) $ by an appropriate subset  of the generators of $ G $, or $ (b) $ by the same set of generators of $ G $, with its gauge parameters being the parameters of $ G $ restricted to a suitable submanifold. In both scenarios, all canonical pairs $ (q^{a},p_{a}) $ of the $ G $-invariant theory are assumed to have well defined transformation laws under the group $ G $. For instance, among the fields $ q^{a} $ one may find gauge fields as well as matter fields; the canonical transformation that will hide the $ G $ symmetry, maps $ (q^{a},p_{a}) $ into $ (Q^{a},P_{a}) $ so that from the $ H $ perspective all $ Q $'s and $ P $'s have well defined transformation laws under $ H $.  For instance, some $ Q $'s transform as gauge fields while the remainder arise in a tensorial representation of $ H $. 

In this paper we have analyzed two systems  that fall into the category $ (a) $ described above; these correspond to pure $ SU(3) $ Yang-Mills  theory, and $ SU(3) $ Yang-Mills theory coupled to a Higgs sector with SSB. In both cases $ G=SU(3) $ and $ H=SU(2) $. The former model allowed us to clarify the meaning of a suitable canonical transformation that lead us to the concept of hidden symmetry --such transformation maps gauge fields of $ SU(3) $ into gauge fields, two doublets and a singlet with respect to the  $ SU(2) $ subgroup. The latter model was useful in order to formulate our notion of hidden symmetry within the context of a well known theory with SSB. The particular scenario of SSB gave an insight into the interpretation of NSGTs; a definite type of these transformations can be seen as the unitary gauge. In both cases, the original symmetry was hidden into the set of SGTs, which we showed corresponds to the $ SU(2) $ group, and the NSGTs, which do not form a group. 

Pure Yang-Mills theory with one compactified UED falls into the category $ (b) $ described above. This theory is formulated to be invariant under the gauge group $SU(N,{\mc M}^{5})$, and the corresponding Poincar\' e group $ISO(1,4)$. Compactification maps the theory into an effective theory invariant under $ SU(N,\mc{M}^{4}) $ and $ ISO(1,3) $. The suitable canonical transformation maps, in this case, gauge fields $ \mc{A}^{a}_{M} $ of $ SU(N,\mc{M}^{5}) $ onto gauge fields $ A^{(0)a}_{\mu} $ and matter fields $ A^{(m)a}_{M} $ of $ SU(N,\mc{M}^{4}) $. As Lie groups $ SU(N,\mc{M}^{5}) $ and $ SU(N,\mc{M}^{4}) $ share the same number of generators, so the map from one to the other cannot involve SSB. However, $ SU(N,\mc{M}^{4}) $ is a subgroup of $ SU(N,\mc{M}^{5}) $ in the following sense: The parameters defining $ SU(N,\mc{M}^{4}) $ are the parameters defining $ SU(N,\mc{M}^{5}) $ restricted to the submanifold $ \mc{M}^{4}$. We conclude after examination of the Lie algebra between SGTs and NSGTs that in the effective theory the SGTs can be identified with the $ SU(N,\mc{M}^{4}) $ group, whereas the NSGTs do not exponentiate into any group. It is important to notice that since there are no broken generators in this scenario, the Higgs mechanism does not operate in the conventional sense; the pseudo-Goldstone bosons needed for this mechanism are provided by an explicit breaking of the Poincar\'e group $ ISO(1,4) $ into $ ISO(1,3) $. Extension of this analysis to theories with more than one compactified UED will be reported elsewhere.

In the Hamiltonian analysis of these models, we found that each canonical transformation translates all the relevant quantities --such as the set the of constraints and the primary Hamiltonian-- from the $ G $ invariant theory to the theory invariant under SGTs and NSGTs. Since each model we analyzed is a first-class constraint system, each canonical transformation maps the gauge generator of the $ G $-symmetry into gauge generators of the SGTs and NSGTs. These results are particularly interesting for the pure $ SU(N,\mc{M}^{5}) $ Yang-Mills theory with one compactified UED and its effective theory; it implies that the gauge structure of the higher-dimensional theory has certainly been rewritten in terms of SGTs and NSGTs. Besides,  by  arguing that the five-dimensional and the effective theory have the same number of physical degrees of freedom, we conclude that the fundamental and the effective theory are equivalent at the classical level.

\acknowledgments{We acknowledge financial support from CONACyT (M\'{e}xico), and J. J. T. also acknowledges
SNI (M\' exico).}

\appendix
\section{Fourier expansion as a canonical transformation}

In this Appendix we will prove that  Fourier expansion is a canonical transformation by showing that it maps conjugate canonical pairs to conjugate canonical pairs. In order to do that,  we will explicitly calculate the nonvanishing Poisson brackets between the zero modes $ (A^{(0)a}_{\mu},\pi^{(0)\mu}_{a}) $, and the $ m $ modes  $(A^{(m)a}_{\mu},\pi^{(m)\mu}_{a}) $ and $(A^{(m)a}_{5},\pi^{(m)5}_{a}) $ as functions of the canonical pairs $ (\mc{A}^{a}_{M},\pi^{M}_{a}) $. We expect to find that this Poisson brackets yield the components of the canonical symplectic two form, proving in this way that the Fourier transformation is indeed a canonical transformation.

We will make use of the following Poisson brackets among the gauge fields and their canonical conjugate momenta, at a fixed time:
\begin{eqnarray*}
\{\mc{A}^{a}_{M}(x,y),\pi^{N}_{b}(x',y')\}_{SU(N,\mc {M})}&=&\delta_{b}^{a}\delta_{M}^{N}\delta (x-x')\delta (y-y')\\
\{\mc{A}^{a}_{M}(x,y),\mc{A}_{N}^{b}(x',y')\}_{SU(N,\mc {M})}&=&\{\pi_{a}^{M}(x,y),\pi^{N}_{b}(x',y')\}_{SU(N,\mc {M})}=0\,,
\end{eqnarray*}
as well as the inverse Fourier transformations
\begin{eqnarray*}
{A}^{(0)a}_{\mu}(x)&=&\frac{1}{\sqrt{R}}\int\ud y\,\mc{A}^{a}_{\mu}(x,y)\ ,\\
{A}^{(m)a}_{\mu}(x)&=&\sqrt{\frac{2}{R}}\int\ud y\,\mc{A}^{a}_{\mu}(x,y)\cos\left(2\pi\frac{my}{R}\right)\ ,\\
{A}^{(m)a}_{5}(x)&=&\sqrt{\frac{2}{R}}\int\ud y\,\mc{A}^{a}_{5}(x,y)\sin\left(2\pi\frac{my}{R}\right)\ ,\\
\pi_{a}^{(0)\mu}(x)&=&\frac{1}{\sqrt{R}}\int\ud y\,\pi_{a}^{\mu}(x,y)\ ,\\
\pi_{a}^{(m)\mu}(x)&=&\sqrt{\frac{2}{R}}\int\ud y\,\pi_{a}^{\mu}(x,y)\cos\left(2\pi\frac{my}{R}\right)\ ,\\
\pi_{a}^{(m)5}(x)&=&\sqrt{\frac{2}{R}}\int\ud y\,\pi_{a}^{5}(x,y)\sin\left(2\pi\frac{my}{R}\right)\ .
\end{eqnarray*}
Also, in order to properly deal with the distributional character of the Poisson brackets, we will use smooth smearing functions $u$ and $v$ defined on ${\mc M}^{4}$.

We proceed to calculate the Poisson bracket between the zero modes with four-dimensional spacetime labels
\begin{eqnarray}
&&\nonumber \{{A}^{(0)a}_{\mu}[u],\pi^{(0)\nu}_{b}[v]\}_{SU(N,{\mc M}^{4})}=\int \ud^{3}x\,\ud^{3}x'\,u(x)v(x')\{{A}^{(0)a}_{\mu}(x),\pi^{(0)\nu}_{b}(x')\}_{SU(N,{\mc M}^{4})}\\
&&\nonumber =\int \ud^{3}x\,\ud^{3}x'\ud y\,\ud y'\,u(x)v(x')\frac{1}{R}\{\mc{A}^{a}_{\mu}(x,y),\pi^{\nu}_{b}(x',y')\}_{SU(N,\mc {M})}\\
&&=\int \ud^{3}x\,\ud^{3}x'\ud y\,\ud y'\,u(x)v(x')\frac{1}{R}\delta_{b}^{a}\delta_{\mu}^{\nu}\delta (x-x')\delta (y-y')=\delta_{b}^{a}\delta_{\mu}^{\nu}[uv]\ \label{A1}.
\end{eqnarray}
The corresponding calculation for the  $m$ modes with four-dimensional spacetime labels reads
\begin{eqnarray}
&&\nonumber \{{A}^{(m)a}_{\mu}[u],\pi^{(n)\nu}_{b}[v]\}_{SU(N,{\mc M}^{4})}=\int \ud^{3}x\,\ud^{3}x'\,u(x)v(x')\{{A}^{(m)a}_{\mu}(x),\pi^{(n)\nu}_{b}(x')\}_{SU(N,{\mc M}^{4})}\\
&&\nonumber =\int \ud^{3}x\,\ud^{3}x'\ud y\,\ud y'\,u(x)v(x')\frac{2}{R}\cos\left( 2\pi\frac{my}{R}\right)\cos\left( 2\pi\frac{ny'}{R}\right)\{\mc{A}^{a}_{\mu}(x,y),\pi^{\nu}_{b}(x',y')\}_{SU(N,\mc {M})}\\
&&=\delta_{b}^{a}\delta_{\mu}^{\nu}\delta ^{mn}[uv]\ \label{A2}.
\end{eqnarray}
Finally, the Poisson bracket between the $m$ modes of the fifth component
\begin{eqnarray}
&&\nonumber \{{A}^{(m)a}_{5}[u],\pi^{(n)5}_{b}[v]\}_{SU(N,{\mc M}^{4})}=\int \ud^{3}x\,\ud^{3}x'\,u(x)v(x')\{{A}^{(m)a}_{5}(x),\pi^{(n)5}_{b}(x')\}_{SU(N,{\mc M}^{4})}\\
&&\nonumber =\int \ud^{3}x\,\ud^{3}x'\ud y\,\ud y'\,u(x)v(x')\frac{2}{R}\sin\left( 2\pi\frac{my}{R}\right)\sin\left( 2\pi\frac{ny'}{R}\right)\{\mc{A}^{a}_{5}(x,y),\pi^{5}_{b}(x',y')\}_{SU(N,\mc {M})}\\
&&=\delta_{b}^{a}\delta ^{mn}[uv]\ \label{A3}.
\end{eqnarray}

As we can see from Eqs. \eqref{A1},  \eqref{A2} and  \eqref{A3}, under the assumption that $(\mc{A}^{a}_{M},\pi^{M}_{a})$ are canonical pairs, one obtains that $({A}^{(0)a}_{\mu},\pi^{(0)\mu}_{a})$, $({A}^{(m)a}_{\mu},\pi^{(m)\mu}_{a})$ and $({A}^{(m)a}_{5},\pi^{(m)5}_{a})$ are canonical pairs.

Conversely, assuming that $({A}^{(0)a}_{\mu},\pi^{(0)\mu}_{a})$, $({A}^{(m)a}_{\mu},\pi^{(m)\mu}_{a})$ and $({A}^{(m)a}_{5},\pi^{(m)5}_{a})$ are canonical pairs, one obtains that $(\mc{A}^{a}_{M},\pi^{M}_{a})$ are canonical pairs. This is achieved by using the Fourier transform and smear functions $u$ and $v$ defined on $\mc M^{5}$, therefore periodic in $y$. These functions will be asked to be even when calculating the Poisson brackets between $\mc{A}^{a}_{\mu}$ and $\pi^{\nu}_{b}$ , so that they can be expanded as follows
\begin{equation}
u(x,y)=\frac{1}{\sqrt{R}} \,u^{(0)}(x)+\sqrt{\frac{2}{R}} \sum ^{\infty}_{m=1}u^{(m)}(x)\cos\left( 2\pi\frac{my}{R}\right)\,;
\end{equation}
and we will demand they be odd when calculating the Poisson brackets between $\mc{A}^{a}_{5}$ and $\pi^{5}_{b}$, and thus expanded as
\begin{equation}
u(x,y)=\sqrt{\frac{2}{R}} \sum ^{\infty}_{m=1}u^{(m)}(x)\sin\left( 2\pi\frac{my}{R}\right)\ .
\end{equation}

In conclusion, from a set of conjugate pairs we obtain, via the Fourier transform, another set of conjugate pairs.

This proof can easily be extended in the presence of more extra dimensions, provided each field has suitable periodic and parity properties on the extra dimensions.


\begin{thebibliography}{99}
%
\bibitem{ATLAS} The ATLAS Collaboration, \textit{Observation of a new particle in the search for the Standard Model Higgs boson with the ATLAS detector at the LHC}, Phys. Lett. \textbf{B716}, 1 (2012).
%
\bibitem{CMS} The CMS Collaboration, \textit{Observation of a new boson at a mass of 125 GeV with the CMS experiment at the LHC}, Phys. Lett. \textbf{B716}, 30 (2012).
%
\bibitem{HM}  F. Englert and R. Brout, Phys. Rev. Lett. \textbf{13}, 321 (1964); Peter W. Higgs, Phys. Rev. Lett. \textbf{13}, 508 (1964); G. S. Guralnik, C. R. Hagen, and T. W. B. Kibble, Phys. Rev. Lett. \textbf{13}, 585 (1964);  Phys. Rev. \tbf{155}, 1554 (1967).
%
\bibitem{GT} Y. Nambu, Phys. Rev. \textbf{117}, 648 (1960); J. Goldstone, Nuovo Cimento \textbf{19}, 154 (1961);
 J. Goldstone, A. Salam and S. Weinberg, Phys. Rev. \textbf{127}, 965 (1962).
 %
 \bibitem{EDA} I. Antoniadis, Phys. Lett. B \textbf{246}, 377 (1990);  N. Arkani-Hamed, S. Dimopoulos and G. R. Dvali, Phys. Lett. B \textbf{429}, 263 (1998);  I. Antoniadis, N. Arkani-Hamed, S. Dimopoulos, and G. R. Dvali, Phys. Lett. B \textbf{436}, 257 (1998).
%
\bibitem{NT} H. Novales-S\' anchez and J. J. Toscano, Phys. Rev. D \textbf{82}, 116012 (2010).
%
\bibitem{NT2} H. Novales-S\' anchez and J. J. Toscano, Phys. Rev. D \textbf{84}, 057901 (2011); \textbf{84}, 076010 (2011).
%
\bibitem{NT3} A. Cordero-Cid, M. G\' omez-Bock, H. Novales-S\' anchez, J. J. Toscano, Pramana J. Phys. \textbf{80}, 369 (2013).
%
\bibitem{EDT} K. R. Dienes, E. Dudas, and T. Gherghetta, Nucl. Phys. \textbf{B537}, 47 (1999); A. M\"uck, A. Pilaftsis, and R. R\"uckl, Phys. Rev. D {\bf 65}, 085037 (2002); N. Uekusa, Phys. Rev. D {\bf 75}, 064014 (2007); R. S. Chivukula, D. A. Dicus, and H.-J. He, Phys. Lett. B {\bf 525}, 175 (2002); S. De Curtis, D. Dominici, and J. R. Pelaez, Phys. Lett. B {\bf 554}, 164 (2003).
%
\bibitem{UEDDM} G. Servant, and T. M. P. Tait, New J. Phys. {\bf 4}, 99 (2002); H. -C. Cheng, J. L. Feng, and K. T. Matchev, Phys. Rev. Lett. {\bf 89}, 211301 (2002); G. Servant, and T. M. P. Tait, Nucl. Phys. \textbf{B650}, 391 (2003); L. Bergstr\"om, T. Bringmann, M. Eriksson, and M. Gustafsson, J. Cosmol. Astropart. Phys. 04 (2005) 004; K. Kong and K. T. Matchev, J. High Energy Phys. 01 (2006) 038; S. Matsumoto and M. Senami, Phys. Lett. B {\bf 633}, 671 (2006); M. Kakizaki, S. Matsumoto, Y. Sato, and M. Senami, Nucl. Phys. {\bf B735}, 84 (2006); M. Kakizaki, S. Matsumoto, and M. Senami, Phys. Rev. D {\bf 74}, 023504 (2006); F. Burnell, and G. D. Kribs, Phys. Rev. D {\bf 73}, 015001 (2006); S. Matsumoto, J. Sato, M. Senami, and M. Yamanaka, Phys. Rev. D {\bf 76}, 043528 (2007); J. A. R. Cembranos, J. L. Feng, and L. E. Strigari, Phys. Rev. D {\bf 75}, 036004 (2007); M. Blennow, H. Melbeus, and T. Ohlsson, J. Cosmol. Astropart. Phys. 01 (2010) 018; J. Bonnevier, H. Melb\'{e}us, A. Merle, and T. Ohlsson, Phys. Rev. D \textbf{85}, 043524 (2012).
%
\bibitem{UEDn} S. Matsumoto, J. Sato, M. Senami, and M. Yamanaka, Phys. Rev. D {\bf 76}, 043528 (2007); S. Matsumoto, J. Sato, M. Senami, and M. Yamanaka, Phys. Lett. B {\bf 647}, 466 (2007); M. Blennow, H. Melb\'{e}us, T. Ohlsson and H. Zhang, J. High Energy Phys. 04 (2011) 052.
%
\bibitem{UEDH} F. J. Petriello, J High Energy Phys. 05 (2002) 003; T. Appelquist and H.-U. Yee, Phys. Rev. D \textbf{67}, 055002 (2003); P. Bandyopadhyay, B. Bhattacherjee and A. Datta, J. High Energy Phys. 03 (2010) 048; M. Blennow, H. Melb\'{e}us, T. Ohlsson and H. Zhang, Phys. Lett. B \textbf{712}, 419 (2012); G. B\'{e}langer, A. Belyaev, M. Brown, M. Kakizaki, and A. Pukhov, Phys. Rev. D \textbf{87}, 016008 (2013).
%
\bibitem{UEDf} A. J. Buras, M. Spranger, and A. Weiler, Nucl. Phys. {\bf B660}, 225 (2003); A. J. Buras, A. Poschenrieder, M. Spranger, and A. Weiler, Nucl. Phys. {\bf B678}, 455 (2004); E. O. Iltan, J. High Energy Phys. 02 (2004) 065; S. Khalil, and R. Mohapatra, Nucl. Phys. {\bf B695}, 313 (2004).
%
\bibitem{UEDc} T. G. Rizzo, Phys. Rev. D {\bf 64}, 095010 (2001); H. -C. Cheng, K. T. Matchev, and M. Schmaltz, Phys. Rev. D {\bf 66}, 056006 (2002); C. Macesanu, C. D. McMullen, and S. Nandi, Phys. Rev. D {\bf 66}, 015009 (2002); G. Bhattacharyya, P. Dey, A. Kundu, A. Raychaudhuri, Phys. Lett. B {\bf 628}, 141 (2005); M. Battaglia, A. Datta, A. De Roeck, K. Kong, K. T. Matchev, J. High Energy Phys. 07 (2005) 033; B. Bhattacherjee, and A. Kundu, Phys. Lett. B {\bf 627}, 137 (2005); {\bf 653}, 300 (2007); D. Hooper, and S. Profumo, Phys. Rep. {\bf 453}, 29 (2007); B. Bhattacherjee, A. Kundu, S. K. Rai, and S. Raychaudhuri, Phys. Rev. D {\bf 78}, 115005 (2008); P. Konar, K. Kong, K. T. Matchev, and M. Perelstein, New J. Phys. {\bf 11} 105004 (2009); S. Matsumoto, J. Sato, M. Senami, and M. Yamanaka, Phys. Rev. D {\bf 80}, 056006 (2009); G. Bhattacharyya, A. Datta, S. K. Majee, A. Raychaudhuri, Nucl. Phys. {\bf B821}, 48 (2009); P. Bandyopadhyay, B. Bhattacherjee, and A. Datta, J High Energy Phys. 03 (2010) 048; B. Bhattacherjee, A. Kundu, S. K. Rai, and S. Raychaudhuri, Phys. Rev. D {\bf 81}, 035021 (2010); D. Choudhury, A. Datta, and K. Ghosh, J High Eenergy Phys. 08 (2010) 051; B. Bhattacherjee, and K. Ghosh, Phys. Rev. D {\bf 83}, 034003 (2011); A. Datta, A. Datta, and S. Poddar, Phys. Lett. B {\bf 712}, 219 (2012); A. Datta, U. K. Dey, A. Shaw, and A. Raychaudhuri, Phys. Rev. D \textbf{87}, 076002 (2013); S. Chang, K. Y. Lee, S. Y. Shim, and J. Song, Phys. Rev. D {\bf 86}, 117503 (2012).
%
\bibitem{NT4} A. Flores-Tlalpa, J. Monta\~{n}o, H. Novales-S\' anchez, F. Ram\'\i rez-Zavaleta, J. J. Toscano, Phys. Rev. D \textbf{83}, 016011 (2011).
%
\bibitem{331} F. Pisano and V. Pleitez, Phys. Rev. D \textbf{46}, 410 (1992); P.H. Frampton, Phys. Rev. Lett. \textbf{69}, 2889 (1992).
%
\bibitem{dirbk64} Dirac, P. A.~M. {\em Lectures on Quantum Mechanics\/}. Belfer Graduate School of Sciences, Yeshiva University, (1964).
%
\bibitem{henbk} Henneaux, M. and Teitelboim, C., \emph {Quantization of Gauge Systems}. Princeton University Press (1992).
%
\bibitem{cas82} L. Castellani, Ann. Phys \tbf{143}, 357 (1982).
%
\bibitem{BRST}C. Becchi, A Rouet, and R. Stora, Commun. Math. Phys.
\textbf{}42, 127 (1975); Ann. Phys. (N.Y.) \textbf{98}, 287 (1976); I.V.
Tyutin, FIAN (P.N: Lebedev Physical Institute of the
USSR Academy of Science), Report No. 39, 1975.
%
\bibitem{AFAB} For a review of the BRST symmetry at both the classical
and quantum levels within the context of the field-antifield
formalism, see J. Gomis, J. Paris, and S. Samuel, Phys.
Rep. 259, 1 (1995).
%
\bibitem{TM} J. Monta\~{n}o, F. Ram\'\i rez-Zavaleta, G. Tavares-Velasco, and J. J. Toscano, Phys. Rev. D \textbf{72}, 055023 (2005);  F. Ram\'\i rez-Zavaleta, G. Tavares-Velasco, and J. J. Toscano, Phys. Rev. D \textbf{75}, 075008 (2007).
%
\bibitem{golbk} Goldstein, H. \emph{Classical Mechanics}. Addison-Wesley Publishing  Company, 2nd edn, (1980).
%
\bibitem{pon05} J. M. Pons, Stud. Hist. Philos. Mod. Phys. \tbf{36}, 491 (2005).
%
\bibitem{hen90} M. Henneaux, C. Teitelboim,  and J. Zanelli, Nucl. Phys. \tbf {B332}, 169 (1990).
%
\end{thebibliography}
\end{document}